\documentclass[final,3p,times]{elsarticle}

\usepackage[colorlinks, linkcolor=red, anchorcolor=blue, citecolor=green]{hyperref}

\usepackage{graphicx}
\usepackage{amssymb}

\usepackage{lineno}

\usepackage[linewidth=1pt]{mdframed}

\usepackage{amsmath,bm}
 
\usepackage{algorithm}
\usepackage{algpseudocode}
\usepackage{amsmath}
\DeclareMathOperator*{\argmax}{argmax}

\usepackage{algorithm}
\usepackage{algorithmicx}
\usepackage{algpseudocode}
\usepackage{amsmath}

\usepackage[utf8]{inputenc}
\usepackage[T1]{fontenc}
\usepackage{nomencl}
\makenomenclature

\biboptions{numbers,sort&compress} 

\usepackage{booktabs}

\journal{Mechanical Systems and Signal Processing}

\begin{document}

\begin{frontmatter}


\title{Payload-agnostic Decoupling and Hybrid Vibration Isolation Control for a Maglev Platform
with Redundant Actuation}


\author{Zhaopei Gong\fnref{label1}}
\author{Liang Ding\corref{cor1}\fnref{label1}}
\ead{liangding@hit.edu.cn}
\cortext[cor1]{Corresponding Author}
\address[label1]{State Key Laboratory of Robotics and System, Harbin Institute of Technology\fnref{label1}}
\author{Shaozhen Li\fnref{label1}}
\author{Honghao Yue\fnref{label1}}
\author{Haibo Gao\fnref{label1}}
\author{Zongquan Deng\fnref{label1}}

\begin{abstract}

Payload-specific vibration control may be suitable for a particular task but lacks generality and transferability required for adapting to the various payload.
Self-decoupling and robust vibration control are the crucial problem to achieve payload-agnostic vibration control.
However, there are problems still unsolved.

In this article, we present a maglev vibration isolation platform (MVIP), which aims to attenuate vibration in payload-agnostic task under dynamic environment.
Since efforts trying to suppress disturbance will encounter inevitable coupling problems, we analyzed the reasons resulting in it and proposed unique and effective solutions.

To achieve payload-agnostic vibration control, we proposed a new control strategy, which is the main contribution of this article.
It consists of self-construct radial basis function neural network inversion (SRBFNNI) decoupling scheme and hybrid adaptive feed-forward internal model control (HAFIMC).
The former one enables the MVIP creating a self inverse model with little prior knowledge and achieving self-decoupling.
For the unique structure of MVIP, the vibration control problem is stated and addressed by the proposed HAFIMC, which utilizes the adaptive part to deal with the periodical disturbance and the internal mode part to deal with the stability.
\end{abstract}

\begin{keyword}
Active vibration control \sep Neural network decoupling \sep Inverse system \sep Adaptive feedforward control \sep Internal mode control


\end{keyword}

\end{frontmatter}


%

\section{Introduction}
\label{S:1}


Space environment provides us a unique condition, on which gravity-related phenomenon such as convection, sedimentation will disappear. 
This environment allows the scientist to make an observation of unique phenomena in many fields.
However, the vibration, which is excited by the periodical or random motion of mechanical parts on space lab, has become a dominant factor impairing experiments' performance \cite{Li2018, Ahadi2013a, Tryggvason2001}.

Various researchers had made efforts to attenuate that effect \cite{LIU201555, Wang2016, Wang2020}.
Some of them are restricted to their physical limits and cannot achieve a satisfying performance (e.g., the piezoelectric actuator only has micron level stroke and is not suitable for low-frequency vibration isolation. Stewart platform transmits vibration by itself structure).
The maglev actuator seems to provide a preferable solution.
Previous researches have demonstrated that the maglev system is capable of achieving non-contact, long-stroke \cite{stabile2017design, Gu2005, Chen2011} and acceptable vibration isolation performance.
Thus we adopt this scheme here and developed the maglev vibration isolation platform (MVIP).

For the foreseeable future, the vibration isolation platform will become an indispensable platform deployed in every space lab.
Different payloads share it at different times, and we call the process of changing payloads as redeployment.
Given the space lab gradually becomes unmanned, that platform should have the ability to update its control parameters based on the perception of the redeployed payload and achieve satisfactory performance autonomously.

However, the MVIP is a strongly coupled nonlinear system with redundant actuation, varying dynamic features caused by redeployed payload and inevitable disturbance.
How to achieve system payload-agnostic decoupling and effective vibration control is the main problem that is going to discuss in this article.

Efforts have been made to maneuver the maglev system. 
Zhang et al.\cite{Zhang2019} proposed a new magnetic negative stiffness isolator based on Maxwell normal stress theory.
It improved the low-frequency vibration isolation performance with high static support stiffness.
\citet{Nguyen2017} presented a new theoretical framework for maglev actuator modeling with simulation and experimental verification.
The actuator's nonlinear characteristic is well stated.
\citet{Sheng2020, Sheng2018} proposed an active damping and disturbance rejection control strategy for a six-axis magnetic levitation stage.
The coupled motion constraints and coupling errors of the magnetic levitation stage are analyzed in detail.
\citet{Jin2019} designed a linear active disturbance rejection controller to attenuate the external disturbances, system parameter perturbations for a five-axis magnetic suspension system.
Many other control theories have been applied on this system, such as adaptive control \cite{Yi2019}, sliding mode control \cite{Gong2017}, and $H_{2}/H_\infty$ control \cite{Chen2019}, etc. 

However, on our way to achieving system payload-agnostic decoupling and vibration attenuation, there is still a gap between the modeling and control.
The control strategy that ignores the system's inherent coupling will not lead to satisfactory performance due to the following reasons,

\begin{enumerate}[{(1)}]
\item 
The dynamic variation and uncertainty have never been considered. The nonlinear actuation and the redeployed payload will significantly change the system's dynamic characteristic and fails the decoupling scheme based on a fixed simplified or pre-known model.

\item The adoption of redundant actuators adds extra input to the coupling system which needs the corresponding constrains to avoid the problem of infinite solutions.
\end{enumerate}

Due to intuitiveness and effectiveness, the inversion system scheme is adopted here. 
It has shown the effectiveness in decoupling and linearization in many nonlinear MIMO systems \cite{Zhang2018}.
Cascade the inversion model with the original system, a pseudo-linear model can be obtained and the system is decoupled. 
However, due to the system non-linearity and varying terms caused by redeployed payloads, the mathematical inversion model could not be obtained precisely. 
The system is hard to achieve precisely and automatically decoupling by using the analytical inversion method. 
Since RBF neural network (RBFNN) has the ability to handle any nonlinear function with unknown parameters in no need of analytical formulation, we adapted it with the system inversion scheme and proposed a new decoupling method, named as the Self-Construct RBFNN Inversion Scheme (SRBFNNI).

The system suffers from the unknown periodical vibration disturbance that comes from different sources. Intuitive is to introduce the disturbance as feedforward, which could be handled by the filtered least mean squared (Fx-LMS) based adaptive feedforward controller \cite{Yi2019, Fallah2019, Ardekani2011, Kuo1999}.
Yi et al.\cite{Yi2019} implemented a conventional adaptive feedforward controller on a hysteretic actuator to achieve active vibration isolation with a promising performance.
Fallah et al.\cite{Fallah2019} designed an adaptive inverse feedforward controller to cancel the regenerative chatter vibrations, which occurs in various machining processes.
However, due to the unique mechanism of MVIP, any noise drift will lead to an unavoidable collision, we developed a hybrid adaptive feedforward internal mode controller (HAFIMC) to maneuver the MVIP, to achieve collision avoidance and vibration isolation simultaneously.
The adaptive feedforward part attenuates the significant periodic vibration, and the internal mode part deals with system stability. 

The paper is organized as follows. 
In Section 2, we derived the system model of the proposed MVIP, including the non-linearity properties of the actuator.
The uncertainty caused by redeployed payloads is analyzed.
In Section 3, the existence of the inversion system is proved. 
Then the proposed SRBFNNI system decoupling scheme is described, derived, and implemented.
The control problem is stated into two parts in our proposed hybrid controller which are separate by adaptive control and internal mode control.
This is described and implemented in Section 4.
The simulation and experiments are carried out in Section 5. 
The result verifies the decoupling performance for a payload-agnostic task, and vibration suppression with external disturbance and conclusion is summarized in Section 6.

\section{System Modeling and Coupling Analysis}
\label{S:2}

\subsection{System Description}
\label{S:2.1}
\begin{figure}[H]
\centering\includegraphics[width=0.75
\linewidth]{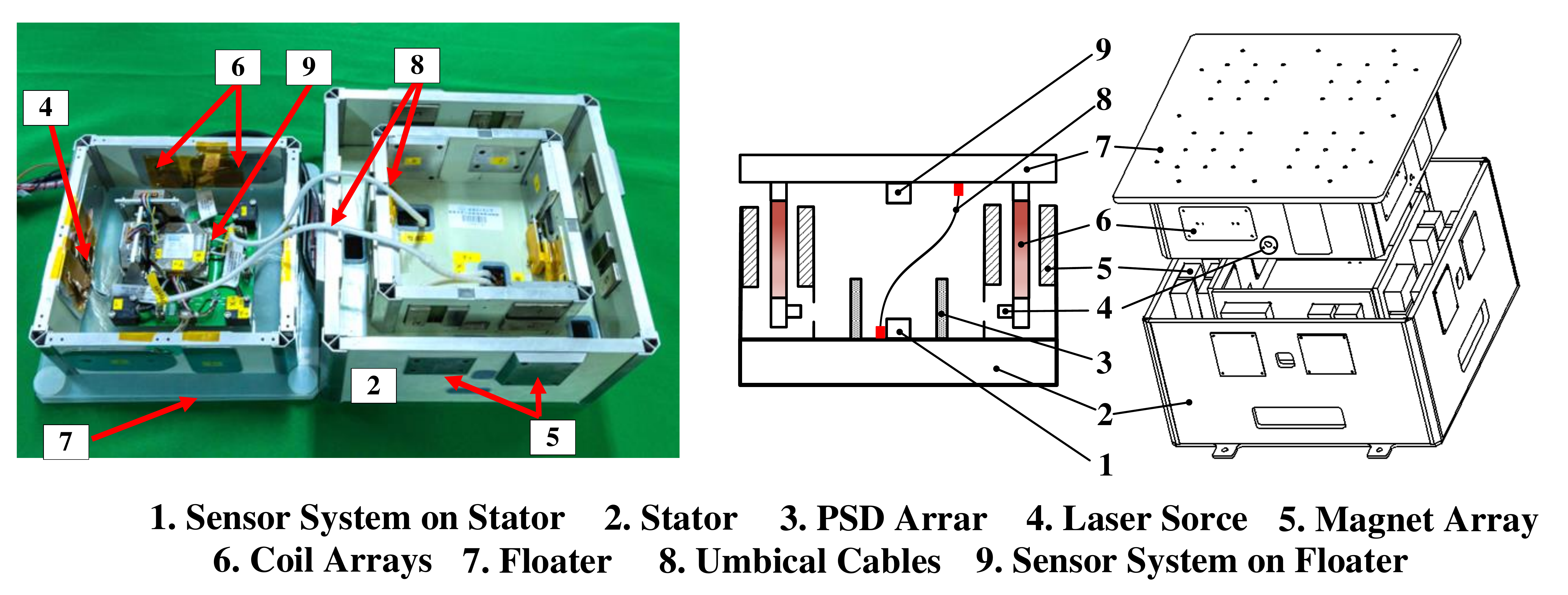}
\caption{Explored view of the MVIP}
\label{fig:systemintro}
\end{figure}

The structure of the developed maglev vibration isolation platform (MVIP) is shown in Fig.\ref{fig:systemintro}. 
The floater has the ability to move in 6 degrees of freedom (DoF), three of rotation and three of translation, within the mechanical closure of the stator.
It is maneuvered directly by eight maglev actuators, which are evenly distributed around the float, with no contact of the stator.
The floater is equipped with six one-axis gyros (three on the floater and three on the stator) and six three-axis accelerometers \cite{Zhu2015}.
The relative position between the floater and the stator is obtained by four two-axis PSD array (Position sensitive device) \cite{Gong2019}.
A mathematical model of the actuator and the motion equation of MVIP will be described in Section \ref{S:2.2}. 
The coupling, non-linearity, and uncertainty features corresponding with the payload-agnostic vibration control will be stated and analyzed in Section \ref{S:2.3}. 

\subsection{Mathematics Modeling of the MVIP}
\label{S:2.2}

\begin{figure}[h]
\centering\includegraphics[width=0.55\linewidth]{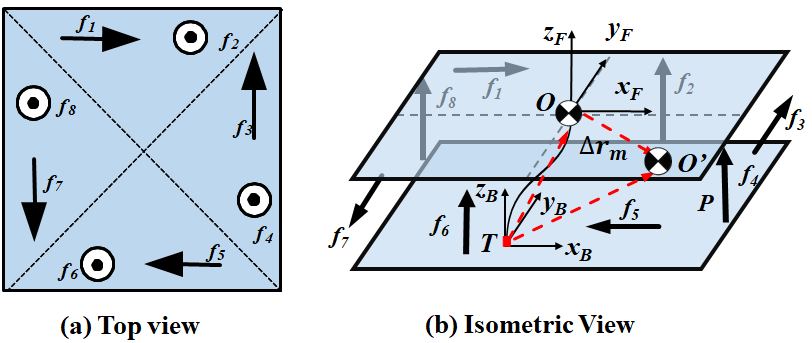}
\caption{Frame definition and actuation scheme}
\label{fig:SystemDynamic}
\end{figure}

The floater is treated as a rigid body and designed to have asymmetric mass distribution.
Thus, its center of geometry (CoG) should coincide with its center of mass (CoM).
However, the payload redeployment significantly shifts CoM from $O$ to $O'$ as shown in Fig.\ref{fig:SystemDynamic}. 

For a given payload, the dynamic equation of MVIP can be deduced as:

\begin{equation}
\label{eq:1}
\left[ \begin{array}{cc}{\boldsymbol{M}} & {-{\boldsymbol{M}}\left(\Delta \boldsymbol{r}_{m}\right)^{ \times}} \\ {\boldsymbol{0}_{3 \times 3}} & {\boldsymbol{{J}_{m}}}\end{array}\right] \ddot{\boldsymbol{X}}\boldsymbol{_{p}}+\left[ \begin{array}{cc}{\boldsymbol{0}_{3 \times 3}} & {-\left(\boldsymbol{\omega} \times \Delta \boldsymbol{r}_{m}\right)^{ \times}} \\ {\boldsymbol{0}_{3 \times 3}} & {-\left(\boldsymbol{J}_{m} \boldsymbol{\omega}\right)^{ \times}}\end{array}\right] \dot{\boldsymbol{X}}\boldsymbol{_{p}}+ \left[ \begin{array}{cc}{\boldsymbol{0}_{3 \times 3}} & {\boldsymbol{M} g} \\ {\boldsymbol{0}_{3 \times 3}} & {g \left(\Delta \boldsymbol{r}_{m}\right)^{ \times}}\end{array}\right] \boldsymbol{X}\boldsymbol{_{p}} + \boldsymbol{D}=\boldsymbol{C}_{M} \cdot \boldsymbol{f}   
\end{equation}
where $\boldsymbol{X_{p}} = [ r_{c x}, r_{c y}, r_{c z}, \theta_{x}, \theta_{y}, \theta_{z}]^{T} $ denotes the motion of floater coordinate ($O$) in stator coordinate ($T$), which is obtained by the PSD sensor array.
$\boldsymbol{M} = diag [ m, m, m]$ and $\boldsymbol{J_{m}} = diag [ J_{x}, J_{y}, J_{z}]$ are the mass and inertia moments matrix of the assembly consisting of the floater and the unknown payload.
$\boldsymbol{\omega} =  [\omega_{x}, \omega_{y}, \omega_{z} ]^{T}$ is the angular velocity of the floater. 
$(\bullet)^{\times}$ is the cross product operator. 
$\boldsymbol{f} = [f_{1} \cdots f_{8} ]^{T}$ is the actuation vector of the eight actuators.
$\boldsymbol{\Delta r_{m}} = [\Delta r_{m x}, \Delta r_{m y}, \Delta r_{m z}]^{T}$ is the shift vector from $O$ to the assembly's CoM ($O'$) .
Since the CoM ($O'$) varies with different redeployed payloads, $\boldsymbol{\Delta r_{m}}$ is difficult to get and regarded as an unknown but unignorable variable. 
$\boldsymbol{C_{M}}$ maps the actuation vector $\boldsymbol{f}$ into wrench in frame $T$,
which can be written as follows:

\begin{small}
\begin{equation}
\label{eq:2}
    C_{M}=\left[ \begin{array}{cccccccc}{1} & {0} & {0} & {0} & {-1} & {0} & {0} & {0} \\ {0} & {0} & {1} & {0} & {0} & {0} & {-1} & {0} \\ {0} & {1} & {0} & {1} & {0} & {1} & {0} & {1} \\ {0} & {y_{m 2}} & 0 & {y_{m 4}} & {0} & {y_{m 6}} & 0 & {y_{m 8}} \\ 0 &  {-x_{m 2}} & {0} &  {-x_{m 4}} & 0 & {-x_{m 6}} & {0} &  {-x_{m 8}} \\ {-y_{m 1}} & {0} & {x_{m 3}}  & {0} & {y_{m 5}} & {0} & -{x_{m 7}} & {0}\end{array}\right]
    \in \mathbb{R}^{6 \times 8}
\end{equation}
\end{small}
in which $(x_{m i}, y_{m i})$ is the location of the $i$-th $(i = 1 \cdots 8)$ actuator in the stator coordinate ($T$) .

\begin{figure}[h]
\centering\includegraphics[width=0.68\linewidth]{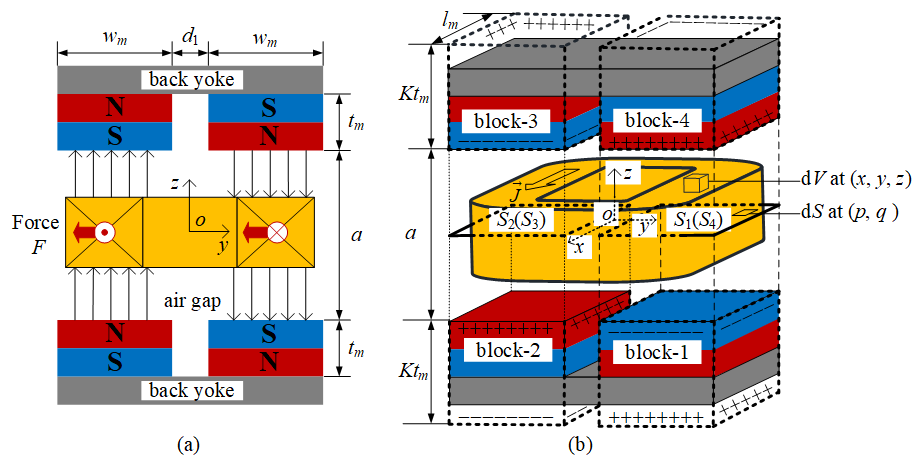}
\caption{Schematics of one maglev actuator unit}
\label{fig:Actuator Model}
\end{figure}

As shown in Fig.\ref{fig:Actuator Model}, the actuation effort is generated by the interaction between the magnet arrays and the winding current. 
Considering the effect of the back yoke, the original magnet block can be equivalent to a block of thickness $K \cdot t_{m}$ ($K$ is thickness coefficient). According to the image theory the electromagnetic actuation effort can be modeled as 

\begin{small}
\begin{equation}
\label{eq:3}
f=\frac{\sigma_{m} J}{4 \pi} \sum_{k=1}^{4} \iiint_{V}\iint_{S_{k}}\left\{\frac{\cos (\langle\vec{J}, \vec{i}\rangle)\left(z-z_{k n}\right)}{\left[(x-p)^{2}+(y-q)^{2}+\left(z-z_{k n}\right)^{2}\right]^{\frac{3}{2}}}-\frac{\cos (\langle\vec{J}, \vec{i}\rangle)\left(z-z_{k p}\right)}{\left[(x-p)^{2}+(y-q)^{2}+\left(z-z_{k p}\right)^{2}\right]^{\frac{3}{2}}}\right\} d S_{k} d V.
\end{equation}
\end{small}
where $f$ is the interaction force between the coil and the magnet arrays.  
$\sigma_{m}=\mu_{0}M$ is the surface magnetic charge density of the equivalent magnet with the magnetization $M(A/m)$ and $\mu_{0}$ is the permeability of free space. 
$J=K_{I} I$  is the amplitude of the current density $(A/m^{2})$ while $K_{I}$ and $I$ are the wire density and the current separately.
$z_{k n}$ and $z_{k p}$ denote the position of the negative and positive magnetic surfaces of $k$-th block separately.

\subsection{Coupling and Non-linearity Analysis}
\label{S:2.3}
The MVIP suffers from the cross-coupling mainly due to three reasons, 1) the actuator's position-dependent nonlinearity, 2) the redundant actuation and most importantly, 3) the system dynamic variation caused by redeployed payload.

\begin{figure}[h]
\centering\includegraphics[width=0.50\linewidth]{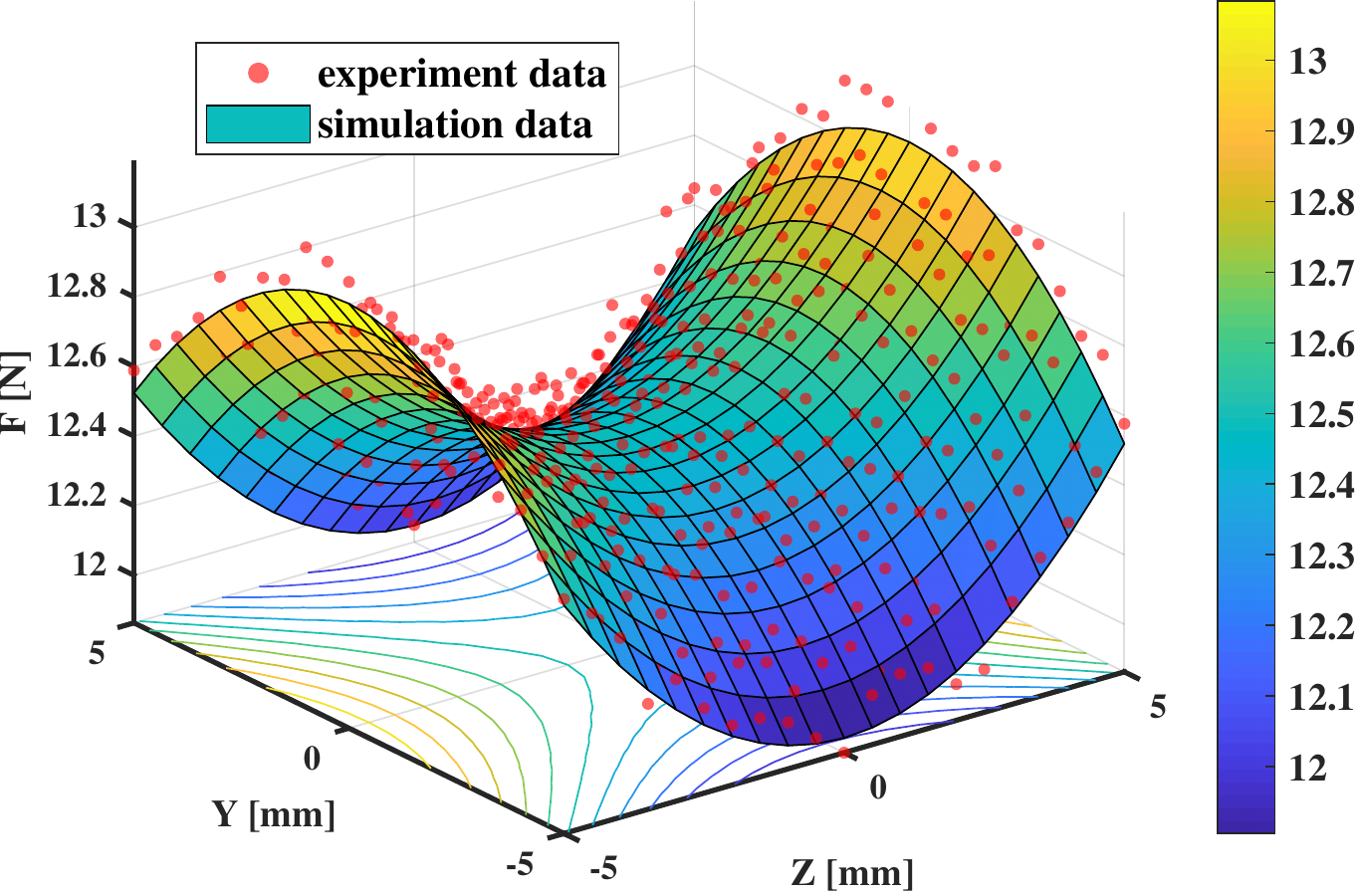}
\caption{The nonlinear position-dependent features of maglev actuator}
\label{fig:4}
\end{figure}

For reason 1), due to the computation expensive of (\ref{eq:3}), it is hard to implement it in the real-time closed loop for feedback compensation.
Experiments and numerical calculation were carried out to obtain the position-dependent features.
As shown in Fig.\ref{fig:4}, the actuator is driven with 1A current and the simulation and experimental obtained data are logged.
It is clearly shown the actuation effort exhibits strong position-dependent properties.
A surface interpolation function $Q$ is adopted to approximate this characteristic as

\begin{equation}
\begin{aligned}
\label{eq:4}
    f &= Q\left(y_{c}, z_{c}\right) \cdot I_{w} + \mathbf{\delta}(\boldsymbol{X}\boldsymbol{_{p}}) \\ &=\left( K_{1} y_{c}^{2} + K_{2} z_{c}^{2} + K_{3} \right) \cdot I_{w} + \mathbf{\delta}(\boldsymbol{X}\boldsymbol{_{p}})
\end{aligned}
\end{equation}
where
$I_{w}$ is the winding current.
$K_{1}$, $K_{2}$ and $K_{3}$ are the approximated current stiffness coefficients of the actuator. $y_{c}$ and $z_{c}$ are the coil's position in the locally actuator coordinate system.
$\mathbf{\delta}$ is the residual unmodeled features function related with system states $\boldsymbol{X}\boldsymbol{_{p}}$.
With this computation cheap method, a feedback compensation could be implemented in real time to deal with the actuation nonlinear problem.

For reason 2), although redundant actuation improves system robustness, it degrades the system to a non-square singular matrix, which has infinite combinations of control solutions.
Most of the solutions may not be executable or could cause a system crash.

We proposed a decoupling scheme based on geometric constraints and energy minimization.
First, we introduced a virtual wrench \begin{small}{$\boldsymbol{\mathcal{F}_{a}} = \boldsymbol{C_M} \cdot \boldsymbol{f} =\left[ F_{N x}, F_{N y}, F_{N z}, \tau_{N x}, \tau_{N y}, \tau_{N z} \right]^{T}$} \end{small} to project the real actuation from a high dimensional space into a low dimensional space. 
Thus, the original system is decomposed into two parts, the non-singular part (from system's states to virtual wrench $\boldsymbol{\mathcal{F}_{a}}$), and the singular part (from virtual wrench $\boldsymbol{\mathcal{F}_{a}}$ to real actuation $\boldsymbol{f}$).
Second, the decoupling problem of the latter part is stated and resolved based on optimization theory as

\begin{subequations}
\label{eq:5}
\begin{equation}
    \left\{\begin{array}{l}{\min \quad \frac{1}{2} \cdot \boldsymbol{f} \cdot (\boldsymbol{Q}^{-1} )^{T} \cdot \boldsymbol{I} \cdot \boldsymbol{Q}^{-1} \cdot \boldsymbol{f}} \\ {\text {s.t.} \quad \ \ \boldsymbol{C}_{M} \cdot \boldsymbol{f} = \boldsymbol{\mathcal{F}_{a}}}\end{array}\right.
\end{equation}
\begin{equation}
    \boldsymbol{f}=\left[ \begin{array}{cc}{\boldsymbol{I}_{8 \times 8}} & {\mathbf{0}_{8 \times 6}}\end{array}\right] \left[ \begin{array}{cc}{(\boldsymbol{Q}^{-1} )^{T} \cdot I \cdot \boldsymbol{Q}^{-1}} & {-\boldsymbol{C}_{M}^{T}} \\ {\boldsymbol{C}_{M}} & {\boldsymbol{0}_{6 \times 6}}\end{array}\right]^{-1} \left[ \begin{array}{c}{\mathbf{0}_{8 \times 1}} \\ \boldsymbol{\mathcal{F}_{a}}\end{array}\right]
\end{equation}
\end{subequations}
where $\boldsymbol{Q} = diag \left[Q_{1}, \cdots , Q_{8}  \right]$ is the surface interpolation function in (\ref{eq:4}).
The algorithm (\ref{eq:5}) also can be extended to a $n$-dimensional input $m$-dimensional output ($n>m$) system, which is a preferable solution for vibration isolation system with redundant actuation.

For reason 3), noticing the $\Delta {r}_{m}$ in (\ref{eq:1}), it couples the system states with each other. 
Taking the translation along the $x$-axis for example, it could be expanded as

\vspace*{-2mm}
\begin{equation}
\label{eq:6}
{\ddot{r}_{c x}=m^{-1}(f_{1} - f_{5} +d) - g \cos{\theta_{x}} + \left( \omega_{y}^{2} + \omega_{z}^{2} \right)\Delta r_{m x} +  \left( \dot{\omega}_{z} - \omega_{x} \omega_{y}\right)\Delta r_{m y} - \left( \dot{\omega}_{y} + \omega_{x} \omega_{z}\right)\Delta r_{m z}}
\end{equation}
\vspace*{-4mm}

From Eq.\ref{eq:6}, we can see that $\ddot{r}_{c x}$ is related with $\boldsymbol{\dot{\theta}} = [\omega_{x}, \omega_{y}, \omega_{z}]^{T}$, $\dot{\omega}_{x}$, $\dot{\omega}_{y}$ and actuation $f_{1}$, $f_{2}$.
This will make any decoupling process based on the ideal assumption fails inevitably.
E.g., under general circumstances, any control effort that applied on the fake CoM ($O$) and expects a translation will be followed by an undesired rotation, which will greatly downgrade the vibration control performance.
The same situation occurs on the other five DoFs.

Based on the above analysis, the MVIP or a class of multi-DoF AVIS (Active Vibration Isolation System) should be regarded as multi-input-multi-output (MIMO), nonlinear, multi-variable cross-coupling system. 
For achieving better vibration isolation performance under dynamic working conditions, more effective linearization and decoupling (L$\And$D) methods should be employed.
\vspace*{-3mm}

\section{System Decoupling with Self-Construct RBFNN based Inversion Scheme}
\label{S:3}

\subsection{Invertibility Analysis of the MVIP System}
\label{S:3.2}
In order to solve the coupling problems, the inversion system is expected to be implemented to achieve L$\And$D.
A brief derivation of the inversion system scheme is given in \ref{Inversion System Scheme}.
Before implementation, the system reversibility must be verified.
Instead of taking winding current, the virtual wrench $\boldsymbol{\mathcal{F}_{a}}$ is selected as system actuation to avoid the singular problem as discussed in Section \ref{S:2.3}.
Then the system input is $\boldsymbol{U} = \left[ u_{x}, \cdots, u_{\theta z }\right]^{T} = \boldsymbol{\mathcal{F}_{a}}$.
Extend system states $\boldsymbol{X}$ to involve $\boldsymbol{\dot{X}_{p}}$, then $\boldsymbol{X} = [ \boldsymbol{X_{p}}, \boldsymbol{\dot{X}_{p}}]=[ r_{c x}, r_{c y}, r_{c z}, \theta_{x}, \theta_{y}, \theta_{z}, \dot{r}_{c x}, \dot{r}_{c y}, \dot{r}_{c z}, {\omega}_{x}, {\omega}_{y}, {\omega}_{z}  ]^{T}$.
Output states $\boldsymbol{Y} = \left [ y_{1}, \cdots, y_{6}\right]^{T} = [r_{c x}, r_{c y}, r_{c z}, \theta_{x}, \theta_{y}, \theta_{z} ]^{T}$.

Ignore the angular velocity multiplication terms and gravity terms (the residual centripetal acceleration in earth orbit is tiny), the MIMO system equation (\ref{eq:7}) can be reorganized as

\begin{small}
\begin{subequations}
\label{eq:7}
\begin{equation}
    \begin{aligned}
    \qquad \qquad \qquad \qquad \boldsymbol{\dot{X}} &= f(\boldsymbol{X}, \boldsymbol{U}) \\
    \left[\begin{array}{l}{\dot{x}_{1}} \\ {\dot{x}_{2}} \\ {\dot{x}_{3}} \\ {\dot{x}_{4}} \\ {\dot{x}_{5}} \\ {\dot{x}_{6}} \\ {\dot{x}_{7}} \\ {\dot{x}_{8}} \\ {\dot{x}_{9}} \\ {\dot{x}_{10}} \\ {\dot{x}_{11}} \\ {\dot{x}_{12}} \end{array} \right] 
    &=\left[\begin{array}{c}
    x_{7} \\ 
    x_{8} \\
    x_{9} \\
    x_{10} \\
    x_{11} \\
    x_{12} \\
    m^{-1} u_{1}  + \dot{x}_{12}\Delta r_{m y} - \dot{x}_{11} \Delta r_{m z} + \delta_{1}(\boldsymbol{X_{P}})\\
    {m}^{-1} u_{2}  -  \dot{x}_{12} \Delta r_{m x} + \dot{x}_{10}\Delta r_{m z} + \delta_{2}(\boldsymbol{X_{P}})\\
    {m}^{-1} u_{3}  + \dot{x}_{11}  \Delta r_{m x} - \dot{x}_{10}\Delta r_{m y} + \delta_{3}(\boldsymbol{X_{P}})\\
    {I_{x}}^{-1}(u_4- u_{2} \Delta r_{mz} + u_{3} \Delta r_{my})+ \delta_{4}(\boldsymbol{X_{P}})
    \\
    {I_{y}}^{-1}
    (u_{5} - u_{3} \Delta r_{m x} + u_{1} \Delta r_{m z} ) + \delta_{5}(\boldsymbol{X_{P}})
    \\
    {I_{z}}^{-1}
    (u_{6} - u_{1} \Delta r_{m y} + u_{2} \Delta r_{m x}) + \delta_{6}(\boldsymbol{X_{P}})
    \end{array}\right] 
    \end{aligned}
\end{equation}
\begin{equation}
\begin{aligned}
    \boldsymbol{Y}&=\boldsymbol{C} \cdot \boldsymbol{X} \\ &= \left[ \boldsymbol{I}_{6 \times 6} \quad \boldsymbol{0}_{6 \times 6} \right] \cdot \boldsymbol{X} 
\end{aligned}
\end{equation}
\end{subequations}
\end{small}

In order to get the mapping from input $\boldsymbol{U}$ to output $\boldsymbol{Y}^{(\alpha)}$ as depicted in (\ref{eq:9}), $\boldsymbol{Y}$ is differentiated until $\boldsymbol{U}$  is explicitly contained in the derivation which could be formulated as

\begin{small}
\begin{equation}
\label{eq:8}
\begin{aligned}
    \boldsymbol{A}(\boldsymbol{U}) &= \left[\ddot{y}_{1}, \ddot{y}_{2}, \ddot{y}_{3}, \ddot{y}_{4}, \ddot{y}_{5}, \ddot{y}_{6} \right]^{T} \\
    &=\left[ \begin{array}{l} 
    m^{-1} u_{1}  + {I_{z}}^{-1} (u_{6} - u_{1} \Delta r_{m y} + u_{2} \Delta r_{m x}) \cdot \Delta r_{m y} - {I_{y}}^{-1}(u_{5} - u_{3} \Delta r_{m x} + u_{1} \Delta r_{m z}) \cdot \Delta r_{m z} + \delta_{1}(\boldsymbol{X_{P}})\\
    m^{-1} u_{2}  - {I_{z}}^{-1}
    (u_{6} - u_{1} \Delta r_{m y} + u_{2} \Delta r_{m x}) \cdot \Delta r_{m x} + {I_{x}}^{-1}(u_4- u_{2} \Delta r_{mz} + u_{3} \Delta r_{my}) \cdot \Delta r_{m z} + \delta_{2}(\boldsymbol{X_{P}})\\
    m^{-1} u_{3} + {I_{y}}^{-1}
    (u_{5} - u_{3} \Delta r_{m x} + u_{1} \Delta r_{m z}) \cdot \Delta r_{m x} - {I_{x}}^{-1}(u_4- u_{2} \Delta r_{mz} + u_{3} \Delta r_{my}) \cdot \Delta r_{m y}+\delta_{3}(\boldsymbol{X_{P}})\\
    {I_{x}}^{-1} (u_4- u_{2} \Delta r_{mz} + u_{3} \Delta r_{my}) + \delta_{4}(\boldsymbol{X_{P}})
    \\
    {I_{y}}^{-1} (u_{5} - u_{3} \Delta r_{m x} + u_{1} \Delta r_{m z})+ \delta_{5}(\boldsymbol{X_{P}})
    \\
    {I_{z}}^{-1} (u_{6} - u_{1} \Delta r_{m y} + u_{2} \Delta r_{m x})+\delta_{6}(\boldsymbol{X_{P}})
    \end{array} \right]
\end{aligned}
\end{equation}
\end{small}

To clarify the relationship between $\boldsymbol{U}$ and $\boldsymbol{Y}$, we take the desire goes back to control in task space by using the Jacobian matrix which could be formulated as

\begin{small}
\begin{equation}
\label{eq:9}
\begin{aligned}
\boldsymbol{J(U)}
    = \frac{\partial \boldsymbol{A}}{\partial \boldsymbol{U}} &= 
    \text{\small $\left( \begin{array}{ccc}
    \frac{\partial \ddot{y}_{1}}{\partial u_{1}} & \ldots & \frac{\partial \ddot{y}_{1}}{\partial u_{6}} \\
    \vdots & \ddots & \vdots \\
    \frac{\partial \ddot{y}_{6}}{\partial u_{1}} & \ldots & \frac{\partial \ddot{y}_{6}}{\partial u_{6}}\\
\end{array} \right)$} \\
&= 
\text{\small $\left(\begin{array}{cccccc} \frac{1}{m}-\frac{{\Delta r_{m z}}^2}{I_{y}}-\frac{{\Delta r_{m y}}^2}{I_{z}} & \frac{\Delta r_{m x}\,\Delta r_{m y}}{I_{z}} & \frac{\Delta r_{mx}\,\Delta r_{mz}}{I_{y}} & 0 & -\frac{\Delta r_{mz}}{I_{y}} & \frac{\Delta r_{ {my}}}{I_{z}}\\ \frac{\Delta r_{ {mx}}\, \Delta r_{ {my}}}{I_{z}} & \frac{1}{m}-\frac{{\Delta r_{ {mz}}}^2}{I_{x}}-\frac{{\Delta r_{ {mx}}}^2}{I_{z}} & \frac{\Delta r_{ {my}}\,\Delta r_{ {mz}}}{I_{x}} & \frac{\Delta r_{ {mz}}}{I_{x}} & 0 & -\frac{\Delta r_{ {mx}}}{I_{z}}\\ \frac{\Delta r_{ {mx}}\,\Delta r_{ {mz}}}{I_{y}} & \frac{\Delta r_{ {my}}\,\Delta r_{ {mz}}}{I_{x}} & \frac{1}{m}-\frac{{\Delta r_{ {my}}}^2}{I_{x}}-\frac{{\Delta r_{ {mx}}}^2}{I_{y}} & -\frac{\Delta r_{ {my}}}{I_{x}} & \frac{\Delta r_{ {mx}}}{I_{y}} & 0\\ 0 & -\frac{\Delta r_{ {mz}}}{I_{x}} & \frac{\Delta r_{ {my}}}{I_{x}} & \frac{1}{I_{x}} & 0 & 0\\ \frac{\Delta r_{ {mz}}}{I_{y}} & 0 & -\frac{\Delta r_{ {mx}}}{I_{y}} & 0 & \frac{1}{I_{y}} & 0\\ -\frac{\Delta r_{ {my}}}{I_{z}} & \frac{\Delta r_{ {mx}}}{I_{z}} & 0 & 0 & 0 & \frac{1}{I_{z}} \end{array}\right)$}
\end{aligned}
\end{equation}
\end{small}

Rank $ \left[\boldsymbol{J(U)} \right] = 6$ and ${\det(\boldsymbol{J(U)})}$ is calculated as
\begin{equation}
\label{eq:10}
    \text{det}(\boldsymbol{J(U)}) = \frac{1}{I_{x}\cdot I_{y} \cdot I_{z}\cdot m^{3}}
\end{equation}

$\text{Det}(\boldsymbol{J(U)})$ will never be zero unless the inertia product or mass of the floater is zero.
So the inequality $\text{det}(\boldsymbol{J(U)}) \neq 0$ is always satisfied in the physical world.
Further, the relative order of MVIP is $\alpha = (2,2,2,2,2,2)$ and it satisfies $\sum_{i=1}^{6} \alpha_{i} = 12 \leq n$, $n$ is the number of system states.
Thus, based on the inversion theory \cite{X.Dai2001}, the original system of MVIP is invertible, i.e., the relationship between system's states and virtual wrench $\boldsymbol{\mathcal{F}_{a}}$ is bi-univocal and unique.
The $\alpha$-th order integral inversion system $\Sigma_{\alpha}$ can be summarized as

\begin{equation}
\label{eq:11}
\begin{aligned}
    \boldsymbol{U} &= \left[u_{x}, u_{y}, u_{z}, u_{\theta x}, u_{\theta y}, u_{\theta z} \right]^{T} = \boldsymbol{\mathcal{F}_{a}}\\
    &= \Sigma_{\alpha} \left(\boldsymbol{X_{p}},  \boldsymbol{\ddot{Y}} \right)
\end{aligned}
\end{equation}

It should be pointed out that due to the decoupling algorithm described in Section \ref{S:2.3} for reason 2), the system is fully actuated now, rather than over actuated or under actuated.
Otherwise, the system will be no longer invertible.

Cascade $\Sigma_{\alpha}$ with the original system (\ref{eq:10}), according to the system inversion theory, an equivalent pseudo-linear system can be obtained.
However, even if $\Sigma_{\alpha}$ can be obtained explicitly, it is challenging to determine the exact parameter under a dynamic working condition for a payload-agnostic task.
A data-driven scheme could be implemented to approximate the inversion system and achieve the task-agnostic self-decoupling.

\subsection{Self-Construct RBFNN Inversion Scheme}
\label{Sec.3.3}

\begin{figure}[h]
\centering\includegraphics[width=0.65\linewidth]{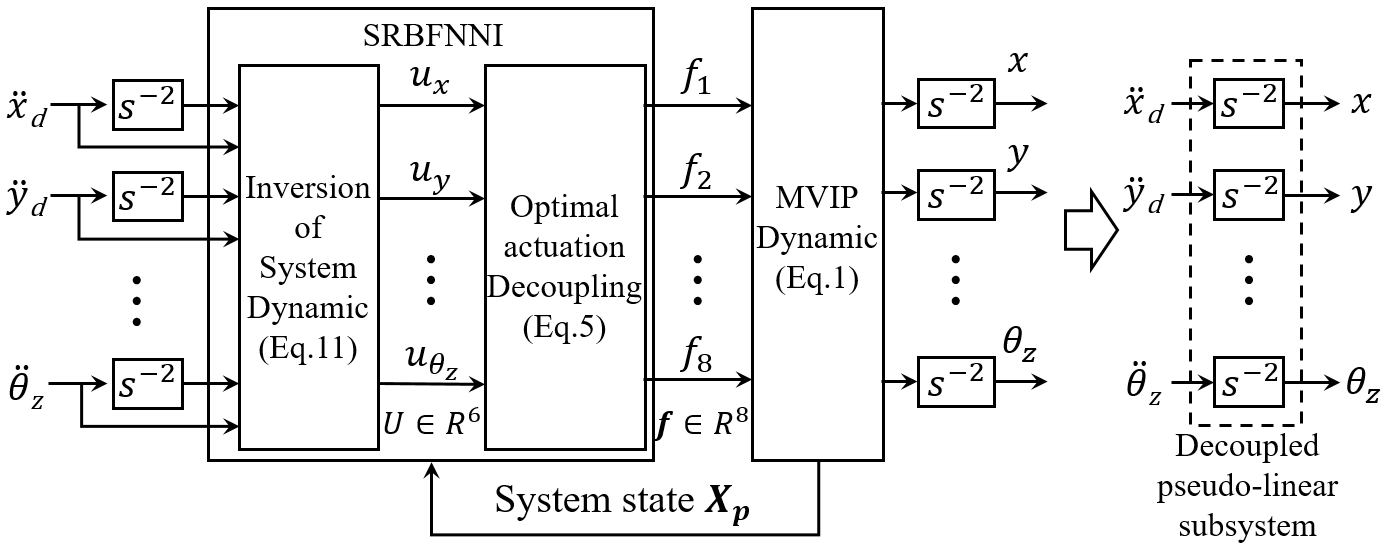}
\caption{Self-Construct Radial Basis Function Neural Networks Inversion Scheme}
\label{fig:SRBFNNI}
\end{figure}

The neural network is a good candidate to have the approximation of the system inversion.
Compared to other machine learning methods such as multilayer perceptron (MLP) and support vector machines (SVM), radial basis function neural networks (RBFNN) have advantages of easy implementation, efficient training and universal approximation \cite{park1991universal,li2018vibration}, which make it suitable for unknown inversion system approximating. 
Given these features, we proposed the Self-Construct RBFNN Inversion (SRBFNNI) scheme as illustrated in Fig.\ref{fig:SRBFNNI}.
An RBF neural network is adopted here to approximate the system inversion $\Sigma_{\alpha}$, in no need of any analytical mathematical model.
It should be noted that the inversion system approached by SRBFNNI, includes not only the system dynamics (\ref{eq:1}) but also the optimal actuation decoupling algorithm (\ref{eq:5}).

Based on (\ref{eq:11}), the inputs of the RBFNN are taken as $\mathbf{x}=[\boldsymbol{X}\boldsymbol{_{p}},  \boldsymbol{\ddot{Y}}] \in \mathbb{R}^{12} $ and output $\mathbf{y}=\boldsymbol{U}=[F_{N x}, F_{N y}, F_{N z}, \tau_{N x}, \tau_{N y}, \\ \tau_{N z}]^{T} \in \mathbb{R}^{6}$.
Cascaded the system inversion $\Sigma_{\alpha}$, which is approximated by a well-trained RBFNN, with the original system (\ref{eq:10}), then the original system can be decoupled into six independent second-order pseudo-linear subsystems.

\begin{figure}[h]
\centering\includegraphics[width=0.36\linewidth]{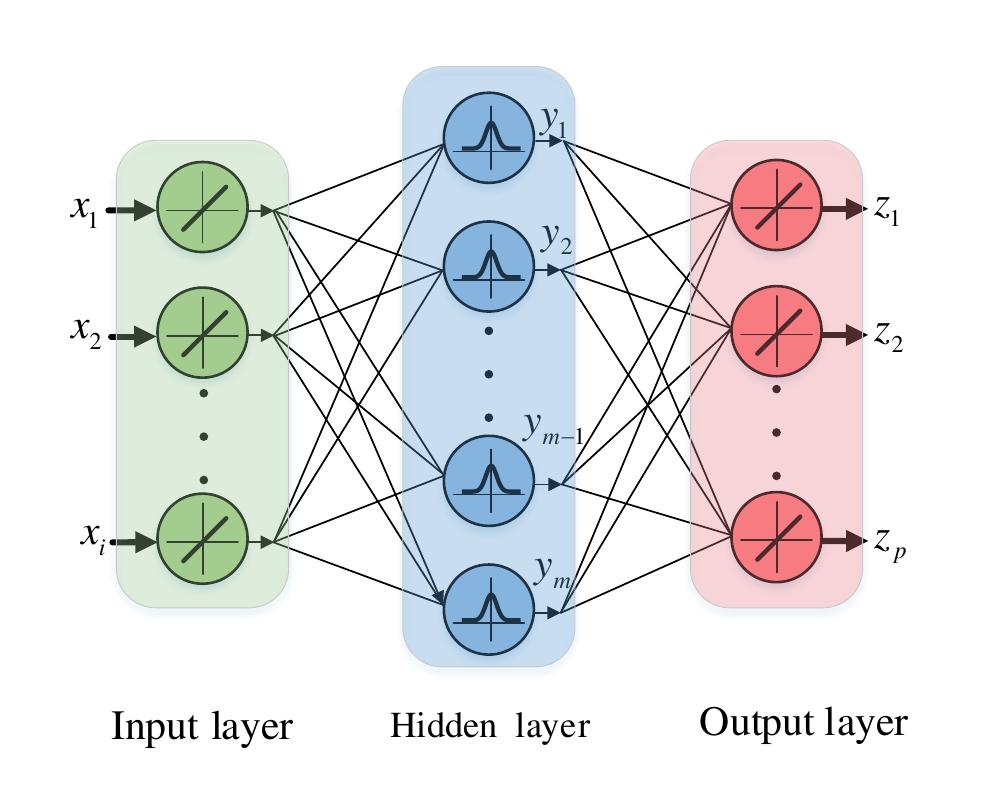}
\caption{Radial Basis Function Neural Networks}
\label{fig:RBFNN}
\end{figure}

As shown in Fig.\ref{fig:RBFNN}, a typical RBFNN consists of a group of processing units referred as neurons or nodes. 
They are organized into three layers:

1) Input Layers:
It consists of $n$ nodes, denoted as ${\mathbf {x}} = \left[x_{1},x_{2},\cdots, x_{n} \right]^{T}$.
Each node is fully connected with the neurons in the next layers.
All the input signals are feed into the next layer simultaneously.

2) Hidden Layers:
Typically, there is only one hidden layer in an RBFNN.
By setting up $p$-dimensional RBF kernel $\varphi (\lVert {\mathbf {x}}-\mathbf {{c}}_{i} \rVert)$ with Euclidean distance between the center of the $i$-th neuron $\mathbf {c}_{i}$ and the input $\mathbf {x}$, the neurons cast the input signal into a high-dimensional feature space. 
The RBF kernel usually takes the form as 

\begin{equation}
    \varphi (\lVert {\mathbf {x}}-\mathbf {{c}}_{i} \rVert) = \text{exp} \left(- \frac{\lVert {\mathbf {x}}-\mathbf {{c}}_{i} \rVert ^{2}}{2 \mathbf{\sigma}_{i}^{2}} \right) \qquad i = 1,2, \cdots, p \\
\end{equation}
where $\sigma_{i}^{2}$ is the radius of this RBF kernel.

3) Output Layers:
It consists of $m$ output nodes, denoted as ${\mathbf {y}} = \left[y_{1},y_{2},\cdots, y_{m} \right]^{T}$.
The output $y_{j}$ is calculated as the weighted sum of outputs from the RBF neurons,i.e.,

\begin{equation}
    y_{j}=\sum_{i=1}^{p} \overline{w}_{i j} \cdot \varphi (\lVert {\mathbf {x}}-\mathbf {{c}}_{i} \rVert) \qquad \qquad j = 1, 2, \cdots, m
\end{equation}
where $\overline{w}_{i j}$ is the weight between the $i$-th RBF neuron and the $j$-th output node.

In order to evaluate the similarity between the trained RBFNN and unknown inversion model, root-mean-square error (RMSE) is adopted.
It is inversely proportional to the similarity between unknown inversion model and RBF neural network.
The desired output $t_{j}$ is compared with the corresponding output $y_{j}$, given a training set of length $Q$, RMSE can be defined as

\begin{equation}
\label{eq:14}
    E_{{RMSE}}=\frac{1}{Q} \sum_{q=1}^{Q} \sqrt{\frac{1}{m} \sum_{j=1}^{m}\left(t_{q j}-y_{q j}\right)^{2}}
\end{equation}

\subsection{Self-Construct RBFNN Inversion Scheme Training and Validation}
\label{Sec.3.4}

\textbf{(1) RBF Neural Network Self-Construct and  Training}

To have a well-trained SRBFNNI, the parameter $p$, $\mathbf {{c}}_{i}$, $\sigma_{i}$ and $\overline{w}_{i j}$ need to be adjusted to minimize $E_{RMSE}$ globally.
We adopt ErrCor \cite{Yu2014} here to create a self-construct RBFNN.
Compared to the LM algorithm \cite{Lian2014} or GD method \cite{8734872}, ErrCor is superior due to its effective and self-construct.
Listed in Algorithm.\ref{alg::conjugateGradient}, we improved the training process to suit multi-output system.
It creates a network iteratively, guaranteeing convergence with the least neurons within only a single one training run and no need for manual adjustment.

\begin{algorithm}[htb]
  \caption{Self-Construct and Training Process of Multi-output RBFNN}
  \label{alg::conjugateGradient}
  \begin{algorithmic}[1]
    \Require
      $\mathbf{x}$: NN input;
      $\mathbf{y}$: NN output;
      $\mathbf{t}$: desired output;
      $S$: maximum iteration;
      $e_{d}$: desired convergence error;
      $N$: maximum number of RBF neurons; 
    \Ensure
      centers of neuron $\mathbf{c}$; radius of neuron $\mathbf{\sigma}$;
      weight $\mathbf{\overline{w}}$;
      number of neuron $p$;
    \State initialize $\mathbf{y}_{1 m}=0$; $p=0$;
    
    \Repeat
      \State record the loop cycle $OutLoopCyc$
      \State compute multi-output error vector  \quad  $E_{q} = \sum_{j=1}^{m} (t_{q j} - y_{q j})^{2}$ for all $q$ in (\ref{eq:14});
      \State localize the max error vector \quad 
      $\beta = \argmax_q E_{q}$;
      \State add a new RBF unit with \quad $\mathbf{c} = \mathbf{x}_{\beta}$,  $\sigma = 1$, $p=p+1$;
      \Repeat
      \State record the loop cycle $InnLoopCyc$
      \State update the $\mathbf{c}$, $\mathbf{\sigma}$ and $\mathbf{\overline{w}}$ with $\boldsymbol{\Delta}_{k+1}=\boldsymbol{\Delta}_{k}-\left(\mathbf{Q}_{k}+\mu_{k} \mathbf{I}\right)^{-1} \mathbf{g}_{k}$ \cite{Hagan1994}
      \Until{($InnLoopCyc \geq S$ or $E_{RMSE} \leq e_{d}$)}
    \Until{($OutLoopCyc \geq N$ or $E_{RMSE} \leq e_{d}$)}
  \end{algorithmic}
\end{algorithm}

Since the MVIP with unknown payload is open-loop unstable, a PID controller is implemented during the sampling process.
After every payload redeployment, we apply position excitation signals to PID controller.
Then the original system input $\begin{small}  \boldsymbol{U} \end{small}$ and  state $\boldsymbol{X_{P}} $ as well as $\boldsymbol{\ddot{Y}}$ are recorded in pairs every 0.002 s for 4 minutes.
Since the value range and units of the raw data are different, they are normalized into $[-1,1]$ to accelerate the training convergence.
With two-thirds of the data set, the algorithm.\ref{alg::conjugateGradient} constructs the SRBFNNI automatically.
The rest data is used for overfitting verification.

\textbf{(2) Excitation Signal Selection}

As stated in Table.\ref{table:nonlin1}, based on same experimental data, two kinds of excitation signal, Random Gaussian Signal (RGS) and Sine Sweep (SS), with three sets of desired $E_{RMSE}$ are tested with Algorithm.\ref{alg::conjugateGradient}.

\begin{table}[ht]
\caption{Effects of excitation Methods and Desired $E_{RMSE}$ on SRBFNNI} 
\centering 
\begin{tabular}{*{5}{c}}
\toprule[1.0pt]
Excitation Signal & Desired  $E_{RMSE}$ & Neuron Numbers & Overfitting \\
\midrule[0.5pt]
RGS & 1e-3 &  6 & No  \\

RGS & 1e-5 &  11 & No  \\

RGS & 1e-7 & 16 & No  \\

SS & 1e-3 &  8 & No  \\

SS & 1e-5 &  14 & No  \\

SS & 1e-7 &  20 & Yes  \\
\bottomrule[1.0 pt]
\end{tabular}
\label{table:nonlin1} 
\end{table}

Since the autocorrelation of RGS is far small than SS, RBFNN trained based on the dataset excited by RGS performs better on approximation and generalization.
A simple data set with high desired $E_{RMSE}$ is more likely to led to overfitting, which happens at the sixth row in Table.\ref{table:nonlin1}.
Considering fast computation and accuracy, the RGS with a desired $E_{RMSE}$ = 1e-5 is adopted here.

\textbf{(3) SRBFNNI Scheme Decoupling Validation}

\begin{figure}[H]
\centering\includegraphics[width=0.45\linewidth]{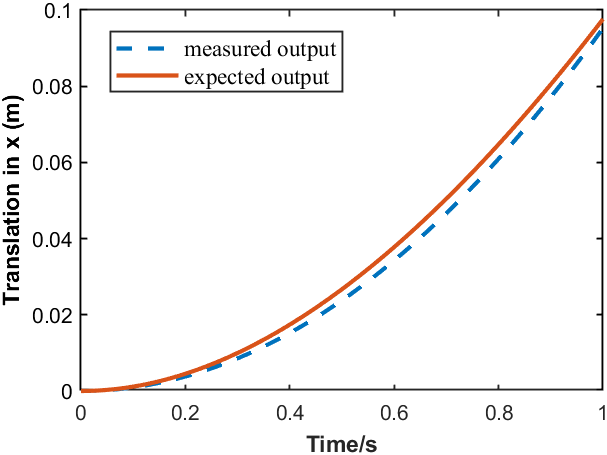}
\caption{SRBFNNI decoupling verification, response of the x-axis with step of 0.2}
\label{fig:7}
\end{figure}

With the self-construct inverse system scheme, we cascade it with the original system.
Since the structure of decoupled system between six channel are the same, we take the pseudo-linear subsystem of x-axis for example to verify the approximation and decoupling effect
It is commanded with the a step signal of 0.2 while other DoF are held by tracking controller.
As shown in Fig.\ref{fig:7}, the expected output is the simulation of the expected model (double integral model, $1/s^2$) excited by the same step input and the measured translation coincides with the expected outputs well.
This results indicates that the proposed SRBFNNI enable the original system to achieve L$\And$D.

\section{Hybrid Adaptive Feedforward Internal Model Control (HAFIMC) for Vibration Isolation}
\label{S:4}

\subsection{Control Objective Statement}
\label{Sec.4.1}

There are two control objectives for MVIP.
One is to minimize the vibration transmitted (through the umbilical cables \cite{zhu2006active}) from the stator to the floater.
The other is to maintain the relative position between the floater and the stator in avoid of collision.
However, they contradict each other.
The position tracking controller will be another vibration transmission pathway itself.
Our solution is achieving them in different frequency bands.
Thus the control objective can be stated as 
\begin{itemize}
	\item Minimize $\left|\ddot{X}_{F}(f) / \ddot{X}_ {B}(f)\right|,  \forall f \in\left[f_{l}, f_{h} \right]$
	\item Force $X_{F}(f) / X_{B}(f) \rightarrow 1, \forall f \in\left[0, f_{l} \right]$
\end{itemize}
where $\ddot{X}_{F}$ and $\ddot{X}_{B}$ are the acceleration of floater and stator, the ${X}_{F}$ and ${X}_{B}$ are the motion correspondingly. $f$, $f_{l}$ and $f_{h} $ are the working, lowest and highest isolation frequencies.

\begin{figure}[H]
\centering\includegraphics[width=0.6\linewidth]{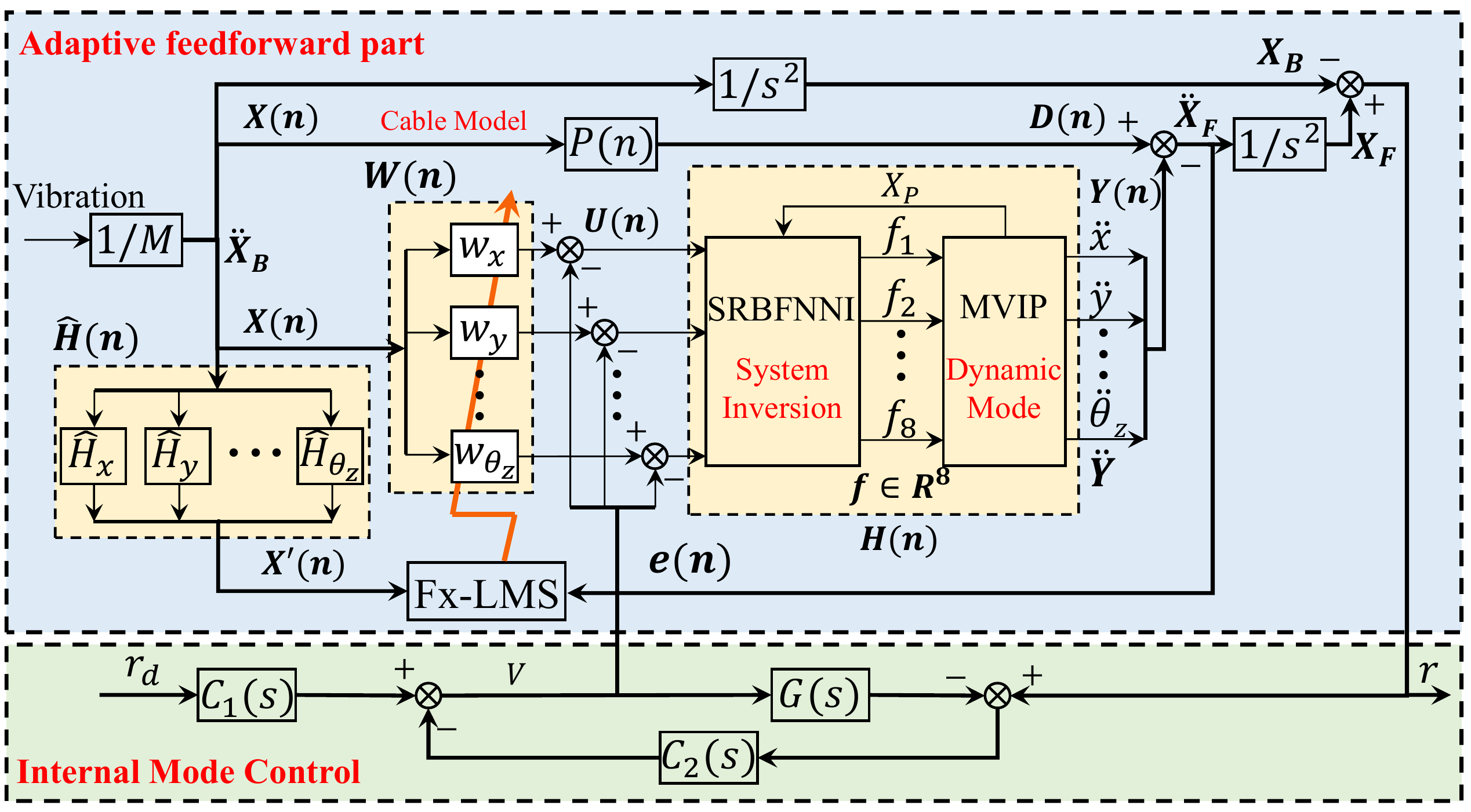}
\caption{hybrid adaptive feed-forward internal model controller}
\label{fig:controller full}
\end{figure}

Compared with low frequency position control \cite{Gong2019},
we proposed an novel hybrid control scheme to achieve these two objectives simultaneously.
As shown in Fig.\ref{fig:controller full}, it consists of two main parts.
The Fx-LMS based adaptive feed-forward part adaptively and actively counteract the vibration to achieve the first objective.
The IMC part maintains the relative position below the isolation frequency in the existence of model uncertainty to achieve the second objective.
This scheme demonstrates superiority over either the single controller when it comes with two control objectives.

\subsection{Adaptive Feedforward Controller Design}
\label{Sec.4.2}

The primary disturbance comes from the stator.
Since we can measure the stator's vibration, an intuitive method is feedforward it on the floater and directly cancel it.
However, we must deal with the unknown transmitting path, which shifts and amplify the vibration.
This is the fundamental motivation for the adaptive feedforward control which can be dealt with the Fx-LMS algorithm \cite{Ardekani2011, Kuo1999}.

As shown in Fig.\ref{fig:controller full}, the adaptive controller utilize the N-order adaptive FIR filter $W_{i}(n)$ and the base vibration $\ddot{X}_{B}$ (also denoted as $X_{i}(n)$) to generate the control output $U_{i}(n)$ and then cancel the vibration transmitted from $\ddot{X}_{B}$.
The cancel error ${e}_{i}(n)$ is the floater's vibration $\ddot{X}_{F}$.
The subscript $i$ denotes the control channel ($i=r_{cx},r_{cy},\dots, \theta{z}$), and the coefficient of $W_{i}(z)$ is denoted as
\vspace*{-5mm}

\begin{align}
    W_{i}(n) =\left[W_{i,0}(n),W_{i,1}(n), \dots,  W_{i,N-1}(n) \right]^{T}
\end{align}
\vspace*{-5mm}

To minimize $\ddot{X}_{F}$, the filter $W_{i}(n)$ is adaptively tuned by the rule as stated in Eq.\ref{eq.update}, which takes two input signals, the reference signal $X^{'}_{i}(n)$ and the error signal $e_{i}(n)$, as performance metrics.
A detailed description is derived in {\ref{FXLMS}} step by step.

\begin{equation}
\label{eq.update}
{W}_{i}(n+1)=\lambda {W}_{i}(n)+2 \mu e_{i}(n) \frac{X_{i}^{'}(n)}{\|{X_{i}^{'}(n)}+p\|^{2}}
\end{equation}
where $\mu$ is the constant value of learning rate, $\lambda$ denotes the forgetting factor which should be close to 1, the minor positive constant $p$ is used to limit the upper value of the weight vector.

It should be pointed out that the reference signal $X_{i}(n)$ and error signal $e_{i}(n)$ comes from the measurement of accelerometer on the stator and floater correspondingly. 
The dynamics model of $H_{i}(n)$ is the combination of SRBFNNI and the original dynamics of MVIP, which forms a pure identical linear transformation and verified in Section.\ref{Sec.3.4}.
$P(n)$ are the umbilical cable mode, which is the main path of stator disturbance.
Due to the difficulties in modeling this vibration transmission path, the adaptive controller is implemented to generate an inverted-phase output to iterative counteract the disturbance transmitted by cables.
Considering the adaptive feedforward loop, the translation function between the floater's acceleration $\ddot{X}_ {F}$  (denoted as $E$) and the stator's acceleration $\ddot{X}_ {B}$ (denoted as $X$) is stated as

\begin{equation}
\label{eq.smc}
    \frac{E(z^{-1})}{X(z^{-1})} = \frac{P(z^{-1})-H(z^{-1})W(z^{-1})}{1+H(z^{-1})}
\end{equation}

It tends to be zero on the condition of Eq.\ref{eq.condition_of_FXLMS} is achieved.
Thus the first objective stated in Section \ref{Sec.4.1} could be achieved.

\subsection{Internal Model Controller Design}
\label{Sec.4.3}

\begin{figure}[h]
\centering\includegraphics[width=0.7\linewidth]{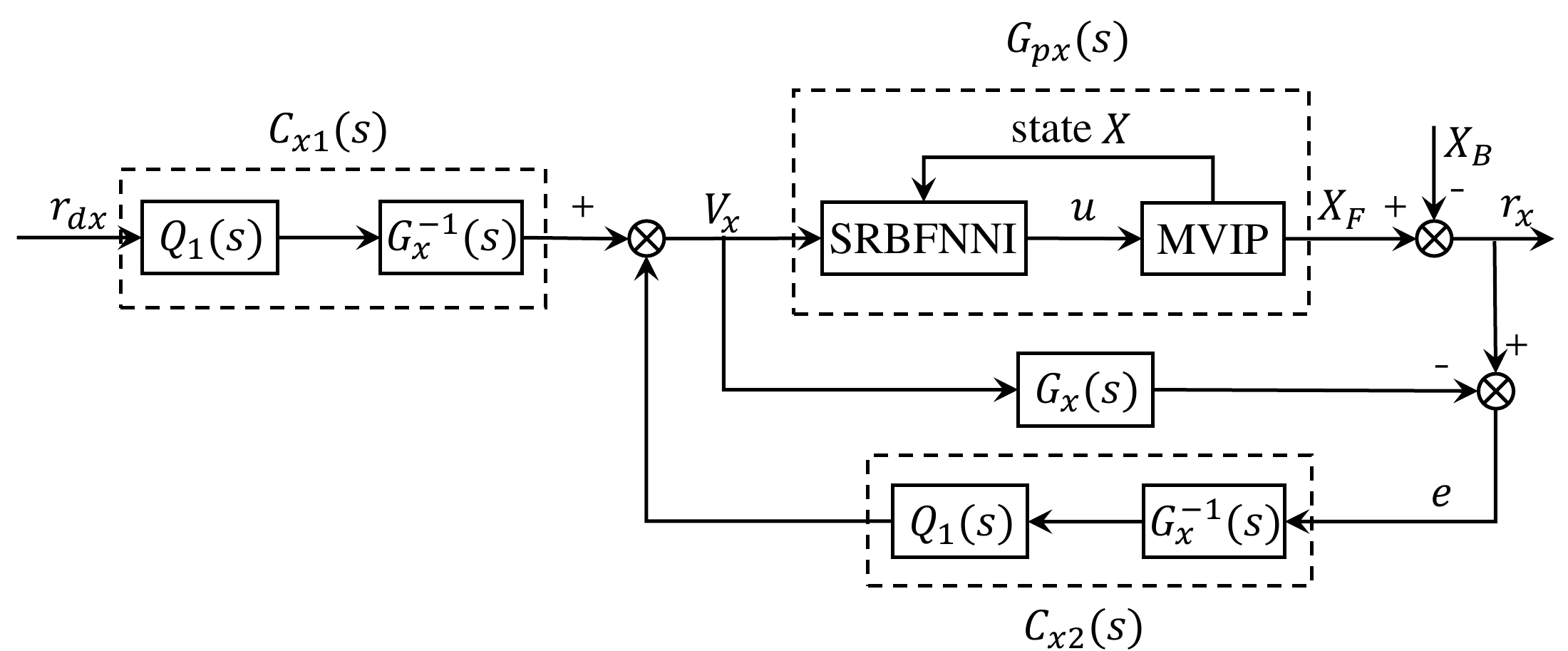}
\caption{Scheme of the 2-DoF internal model controller}
\label{fig:IMC controller}
\end{figure}

Considering the modeling error and the inevitable signal noise, the 2-DoF internal model control (IMC) is a good candidate for achieving system stability and relative position maintain performance. 
Taking the $x-$translation channel for example, the 2-DoF IMC controller can be simplified as shown in Fig.\ref{fig:IMC controller}.

Compared with common implementation, there are some difference in our scheme.
The input of the IMC $r_{dx}$ is the desired relative position.
The output of the IMC is the real relative position, denoted as $r_{x} = X_{F} - X_{B}$.
Although the absolute motion $X_{F}$ and $X_{B}$ is not measurable, $r_{x}$ could be obtained with the PSD sensor array.

The $G_{p x}$ is the real plant model of the $x-$channel.
The $G_{x}$ is the nominal transfer function of the decoupled pseudo-linear subsystem, which can be denoted as

\begin{equation}
\label{eq:simple model}
    G_{x}(s) = G_{p x}(s) - \Delta G_{x}(s) = 1/s^{2}
\end{equation}

$C_{x 1}$ and the $C_{x 2}$ are the corresponding internal mode controller.
$V_{x}$ is the control input and $e$ is the error.
Then the system output can be calculated as,

\begin{equation}
    r_{x}=\frac{G_{p x}(s) C_{x 1}(s)}{1+C_{x2}(s)\left[G_{px}(s)-G_{x}(s)\right]}r_{d x}(s)-\frac{1-G_{x}(s) C_{x 2}(s)}{1+C_{x 2}(s)\left[G_{px}(s)-G_{x}(s)\right]}X_{B}(s)
\end{equation}

When the nominal model is accurate ($G_{p x}(s) = G_{x}(s)$), system output can be formulated as

\begin{equation}
     r_{x} = G_{p x}(s) C_{x 1}(s) r_{d x}(s) - (1-G_{x}(s) C_{x 2}(s)) X_{B}(s)
\end{equation}

It is clearly shown that the relative position maintain performance relays on $C_{x 2}$ and the command tracking performance relays on $C_{x 1}$.
It should be noted that, the relative command is always set as constant 0 for most time and the relative position maintain performance coincides exactly with the second control objective stated in Section.\ref{Sec.4.1}.
Since these two performance is related with frequency, we introduce two filters $Q_{1}$ and $Q_{2}$ as

\begin{equation}
\label{eq.filter}
    \left\{\begin{array}{l}{C_{x 1}(s)= Q_{1}(s)/G_{x}(s)} \\ {C_{x 2}(s)= Q_{2}(s)/G_{x}(s)}\end{array}\right.
\end{equation}

\begin{figure}[h]
\centering\includegraphics[width=0.51\linewidth]{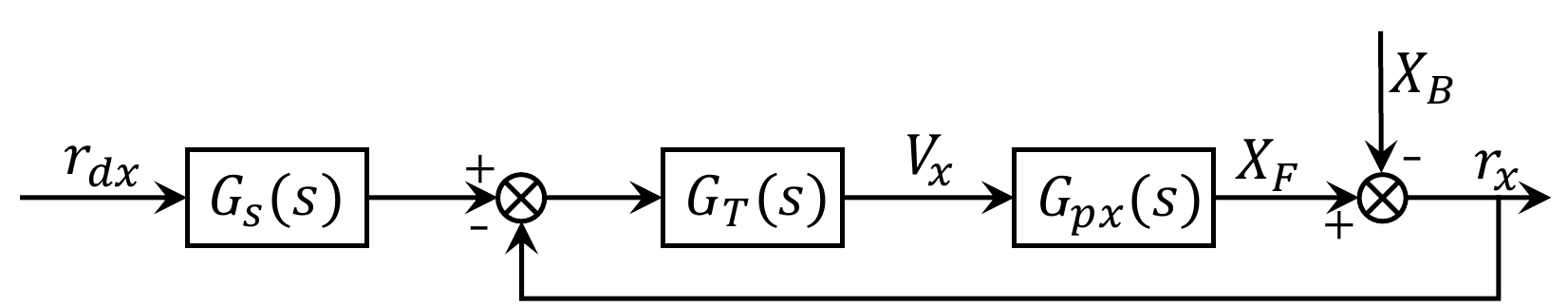}
\caption{Structure of the modified 2-DoF controller}
\label{fig:Modified IMC controller}
\end{figure}

Further, the modified IMC control scheme is illustrated in Fig.\ref{fig:Modified IMC controller}, with

\begin{equation}
\label{eq.simplified_IMC}
    \left\{\begin{array}{l}{G_{s}(s) =C_{x 2}^{-1}(s)} C_{x 1}(s) \\ {G_{T}(s) = [1-G_{x}(s) C_{x 2}(s)]^{-1} } C_{x 2}(s)\end{array}\right.
\end{equation}

To design $Q_{1}(s)$ and $Q_{2}(s)$, we define the error transfer function as

\begin{equation}
\label{eq.error}
    \begin{aligned} E(s) &=r_{d x}(s)-r_{x}(s) \\ &=\frac{\left(G_{T}(s) G_{p x}(s)(1-G_{s}(s)) + 1\right) }{1+G_{T}(s) G_{px}(s)} r_{d x}(s) +  \frac{1}{1+G_{T}(s) G_{px}(s)} X_{B}(s)\end{aligned}
\end{equation}

The relative position maintain performance ($E(s)/X_{B}(s)$) and the command tracking performance ($E(s)/r_{dx}(s)$) can be tuned individually.
For the former, making $r_{dx}(s) = 0$, $G_{px}(s) = G_{x}(s)$ and (\ref{eq.error}) can be rewritten as

\begin{equation}
    E(s)/X_{B}(s) = 1-Q_{2}(s)
\end{equation}

To achieve the second control objective, $Q_{2}(j\omega)$ should be an low-pass filter with a cut-off frequency of $\varepsilon_{2}$ as formulated in Eq.\ref{eq.q2}.
The larger $\varepsilon_{2}$, the higher frequency $X_{F}$ can track $X_{B}$.
Setting $\varepsilon_{2} =2 \pi f_{L}$.
When $\omega < 2\pi f_{l}$, $E(j\omega)$ tends to be zeros, which means the floater tends to follows the stator.
When $\omega > 2\pi f_{l}$, $E(j\omega)$ tends to be $X_{B}(j\omega)$, which means the floater tends to keeps steady. 
Thus the second objective mentioned in Section \ref{Sec.4.1} is achieved.

\begin{equation}
\label{eq.q2}
    Q_{2}(s) = \varepsilon_{2}^{2}/(s + \varepsilon_{2})^{2}
\end{equation}

For the command tracking, making $X_{B}(s) = 0$, $G_{px}(s) = G_{x}(s)$ and we can get

\begin{equation}
    E(s)/r_{dx}(s) = 1-Q_{1}(s)
\end{equation}

To realize the fast convergence, the $Q_{1}$ could be designed as a low-pass filter as

\begin{equation}
\label{eq.q1}
    Q_{1}(s) = \varepsilon_{1}^{2}/(s + \varepsilon_{1})^{2}
\end{equation}

The larger the $\varepsilon_{1}$, the better command tracking performance.
Since the $r_{d x}$ is always set to be 0, we set the $\varepsilon_{1} = 5f_{l}$ to ensure fast convergence and avoid command noise.
According to (\ref{eq.q2}) and (\ref{eq.q1}), in different mode, $E(s)$ are all two-order system with high-pass characteristics.
Thus it can be concluded that the floater can track the stator's step and sinusoidal motion below $f_{l}$ and track the constant zero command with no steady-state errors.

According to the IMC control theory, the stabilization condition for this system with any $\omega$ can be obtained as

\begin{equation}
\label{eq.stable}
    \left|G_{T}(j \omega) G_{x}(j \omega)\right| \overline{l}_{m}<1
\end{equation}
where $\overline{l}_{m}$ is the upper bound of $\Delta G_{x}$.

Substituting (\ref{eq:simple model}) and  (\ref{eq.filter}) into (\ref{eq.stable}) we can have

\begin{equation}
\left|\left(\frac{1}{\varepsilon_{2}} s + 1\right)^{2} - 1\right|>\overline{l}_{m}
\end{equation}

It is clear shown that for certain boundary limit $\overline{l}_{m}$ , the system can be stabilized by selecting the $\varepsilon_{2}$.

\section{Experimental Results}
\subsection{Experimental System Setup}

\begin{figure}[H]
\centering\includegraphics[width=0.95\linewidth]{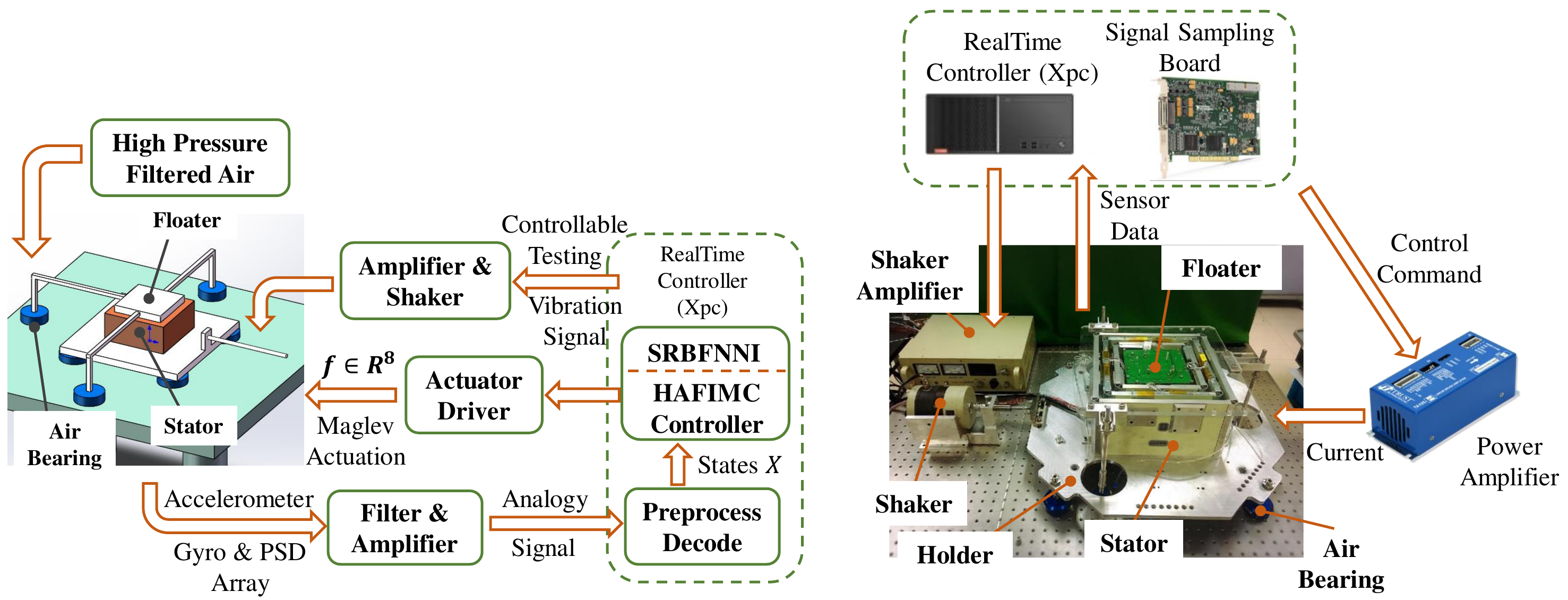}
\caption{Configuration of the MVIP experiment system and the hardware platform}
\label{fig:11}
\end{figure}

To verify the decoupling and the vibration isolation performance with our proposed scheme, we build an experimental vibration system.
As shown in Fig.\ref{fig:11}, it consists of MVIP, air bearing supporters and controllable shaker.
The structure and the equipped sensors of MVIP are described in Section.\ref{S:2.1}.

The sensors' outputs are filtered and amplified by our homemade processing board first, then sampled and processed by a commercial signal sampling board of 18-bits resolution.
A desktop computer with i7 core running  Simullink-Realtime OS is set up here to be the host real-time controller (xPC).
The decoding algorithm is running at 5 kHz on xPC to convert the sensor data into system states.
Then the calculated system states are feed into the proposed SRBFNNI and the HAFIMC which are running at 2 kHz in xPC.
The eight actuators are driven by linear amplifiers (TA115, Trust Automation Inc).
Based on this system, we carried out several experiments as described from Sections \ref{S.5.2} - \ref{S.5.4}, the parameter and coefficient of the controller is listed in Table.\ref{table:2}.

\begin{table}[ht]
\caption{Parameter of proposed controller} 
\centering 
\begin{tabular}{*{4}{c}}
\toprule[1.0pt]
Parameter & Value & Parameter & Value \\
\midrule[0.5pt]
Lowest isolation frequency $f_{l}$ [$Hz$] & 10 &  Learning rate $\mu$ & 0.004  \\
Highest isolation frequency $f_{h}$ [$Hz$] & 300 &  Anti-saturation coefficient $p$ & 0.001  \\
Adaptive Filter Length $l_{w}$ & 65 & forgetting factor $\lambda$ & 0.998  \\
First coefficient of IMC $\varepsilon_{1}$  & 31.4 & Second coefficient of IMC $\varepsilon_{2}$ & 69.08  \\

\bottomrule[1.0 pt]
\end{tabular}
\label{table:2} 
\end{table}

To attenuation and avoid resonance in MVIP, we analyze the floater's nature frequency with finite element analysis (FEA).
The results indicate its first-order resonance is 269.4 Hz and 286.3 Hz for the second-order.
A second-order low-pass-filter with a cut-off frequency of 240 Hz is added in the actuator's power amplifiers and verified by a frequency response analyzer (FRA).
A band-pass filter with a center frequency of 268 Hz is also added to the accelerometer signal acquisition channel. 
Since the umbilical cables with small stiffness are the only physical connection between floater and stator, the system resonant is below 0.2 Hz and can be neglected.

To create a vibrating environment, we build an air floating experimental vibration system.
The stator and floater of MVIP are levitated by air bearings individually. 
The stator sits on an aluminum plate supporter levitated by four air bearings located below the plate, and three frame holder levitates the floater with air bearings under it.
With high-pressure air, the floater and stator can move freely and individually on the surface with almost no friction.
Then the stator can be excited along the glass surface by a linear shaker controlled by xPC. 
With this system, we can verify our scheme in three DoF for each experiment and all six DoF for different deployment.

\subsection{SRBFNNI Decoupling Properties Verification}
\label{S.5.2}

\begin{figure}[H]
\centering\includegraphics[width=0.65\linewidth]{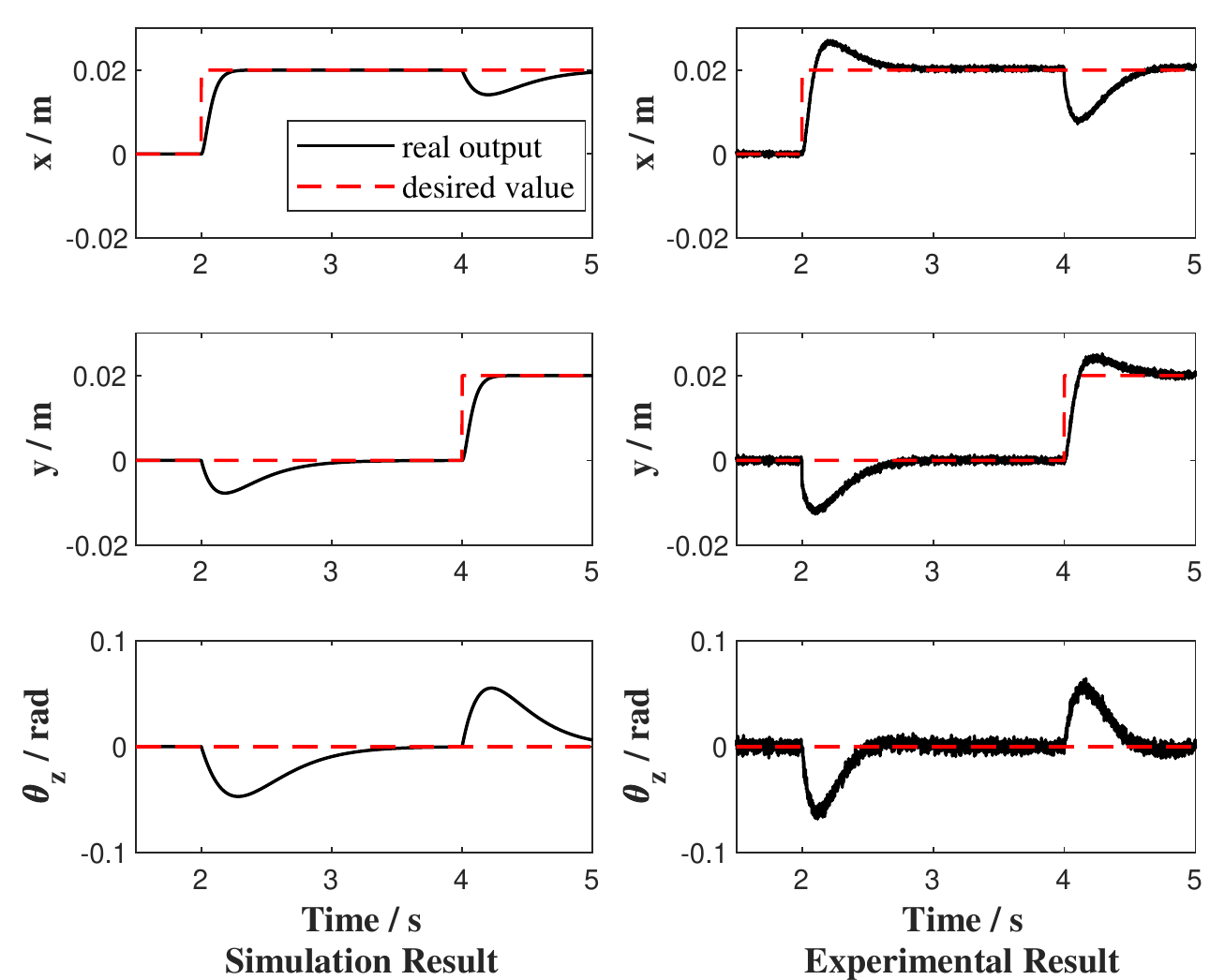}
\caption{System cross-coupling properties without SRBFNNI decoupling}
\label{fig:12}
\end{figure}

\begin{figure}[H]
\centering\includegraphics[width=0.65\linewidth]{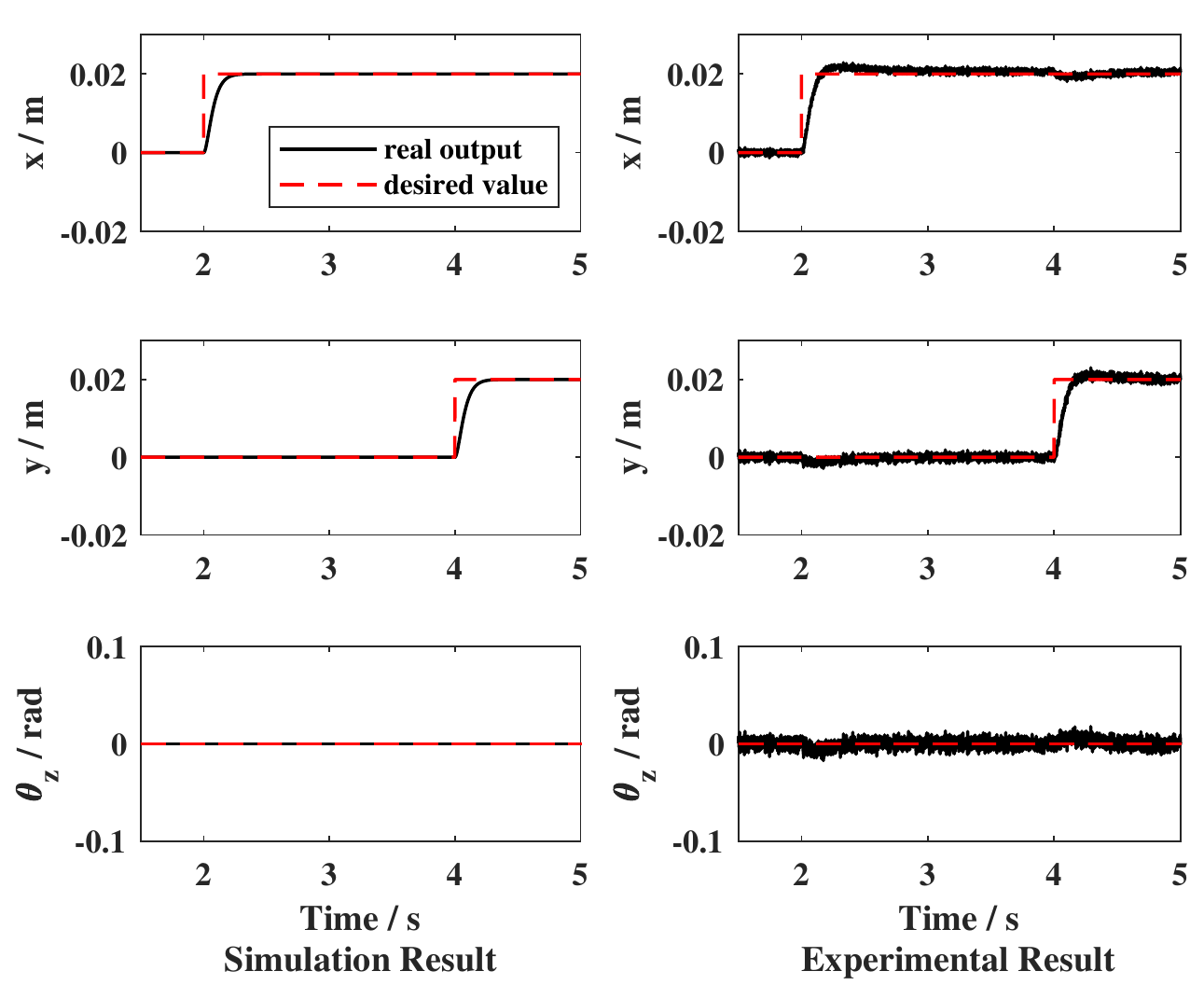}
\caption{System cross-coupling properties with SRBFNNI decoupling}
\label{fig:13}
\end{figure}

Experiments and simulations are carried out to verify the effectiveness of the SRBFNNI decoupling scheme.
With the help of air floating system, MVIP is maneuvered by internal mode controller whose parameters are listed in Table.\ref{table:3} in 3-axis (translation with the $x$ and $y$-axis and rotation with the $z$-axis).
An experimental payload around 5 kg is placed on the floater in a non-central position to mimic the payload redeployment.
Then the $x$-axis relative position signal steps from 0 to 0.02 m at $t$ = 2 s, and the $y$-axis relative position signal steps from 0 to 0.02 m at $t$ = 4 s.
In the case of without decoupling and with SRBFNNI decoupling, the simulation and experimental results are shown in Fig.\ref{fig:12} and Fig.\ref{fig:13}, respectively.

Fig.\ref{fig:12} shows that without SRBFNNI, since the desired system dynamic is different from the real plant, whether the step of the $x$-axis or the step of the $y$-axis, they all caused apparent coupling and shift in the other two channels.
The steps of the $x$-axis caused 0.008 m overshoot in the $y$-axis and 0.04 rad in the $\theta_{z}$ at 2 s.
The similar happens in $y$-axis.
The simulation results are basically consistent with the experimental results, except that the experiment results are affected by the unavoidable noise and the uncertain mass parameters.
The overshoot at 2 s of the $x$-axis and the overshoot at 4 s of the $y$-axis in experiment results is mainly ascribed to the inaccurate mass estimation.
It can be seen that after a payload redeployment, the system dynamics changed a lot, and the coupling effect between different channels is so apparent that it cannot be ignored.

Instead, Fig.\ref{fig:13} shows that in simulation case, with the SRBFNNI decoupling scheme, the step signal of the $x$-axis or the $y$-axis made almost no disturbance to other channels, and in experiment case, the disturbance is so slight that can be ignored.
This result shows that with the help of SRBFNNI, the nonlinear MVIP system with unknown dynamics can be decoupled besides satisfactory performance.
Further, the overshoot caused by parameter uncertain is effectively suppressed, and the control precision is improved.

\subsection{Active Vibration Isolation Control with Internal Mode Controller}
\label{S.5.3}

\begin{figure}[H]
\centering\includegraphics[width=0.8\linewidth]{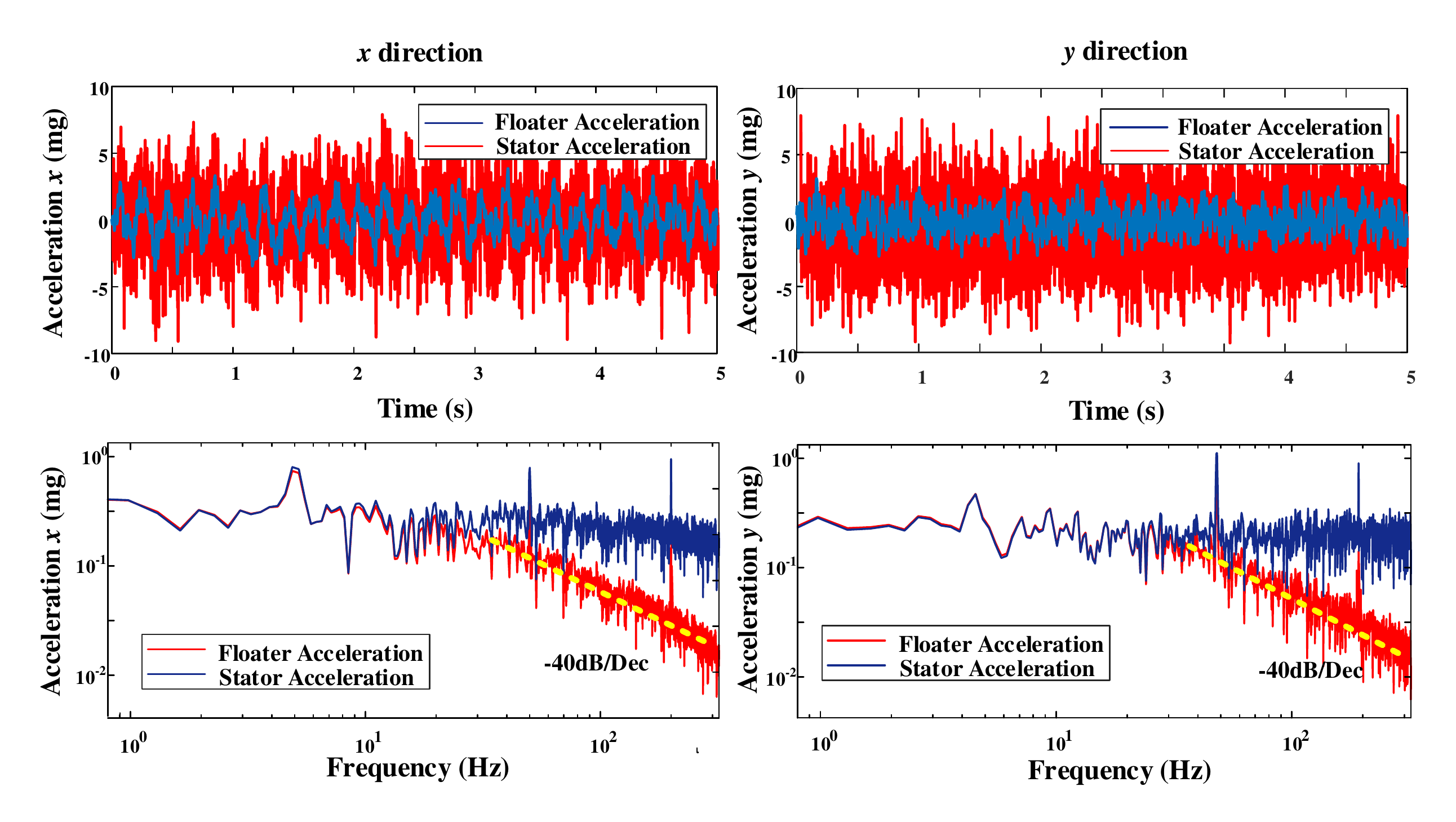}
\caption{Comparison of vibration level on stator and floater with IMC control}
\label{fig:14}
\end{figure}

Since the system is decoupled, experiments are further carried out to verify the achievement of the control objectives.
Taking the random vibration as the background noise, and 5 Hz, 50 Hz and 200 Hz periodic vibrations as low, medium and high frequencies noise respectively, the shaker excite the stator along the $x$-axis and the $y$-axis individually.
Translation along the $x, y$-axis and rotation with the $z$-axis of MVIP are controlled by internal mode controller without feedforward part.
The acceleration of stator and floater in time domain and frequency domain is logged and presented in Fig.\ref{fig:14} individually.

Since the $\varepsilon_{2}$ is set as 69.08 rad/s, the floater should track the vibration below that frequency and keep stead above it.
As it is shown in the lower part in Fig.\ref{fig:14}, the floater tracks the 5 Hz signal well and exhibits a -40 dB/Dec attenuation of random vibrations above 10 Hz which can be attributed to the design of filter $Q_{2}$.
This also coincidence with the time domain results. For the $x$-axis, the acceleration RMSE values of stator and floater are 2.68 mg and 1.05 mg correspondingly. For $y$-axis, the values are 2.81 mg and 1.43 mg.
The results show that the first control objective is achieved.
However, due to the existence of cable transmitting vibration from stator to floater, the suppression for medium and high frequency periodical vibration is not satisfied as listed in Table.\ref{table:3}.
For the $x$-axis, the 50 Hz disturbance signal is only attenuated by -10.86 dB, and for the 200 Hz case, it is only attenuated by -21.49 dB, which does not coincide with the decay ratio.

\begin{table}[H]
\caption{Vibration isolation performance in different control mode} 
\centering 
\begin{tabular}{*{5}{c}}
\toprule[1.0pt]
Frequency & Control Mode & X-Axis [$dB$] & Y-Axis [$dB$] \\
\midrule[0.5pt]
$5 Hz$ & IMC &  -1.02 & -0.84 \\
$5 Hz$ & HAFIMC &  -2.36 & -1.93 \\
$50 Hz$ & IMC & -10.86 & -11.43 \\
$50 Hz$ & HAFIMC & -22.51 & -20.42 \\
$200 Hz$ & IMC &  -21.49 & -20.71 \\
$200 Hz$ & HAFIMC &  -39.75 & -37.61 \\

\bottomrule[1.0 pt]
\end{tabular}
\label{table:3} 
\end{table}

\subsection{Active Vibration Isolation Control with HAFIMC}
\label{S.5.4}

\begin{figure}[H]
\centering\includegraphics[width=0.80\linewidth]{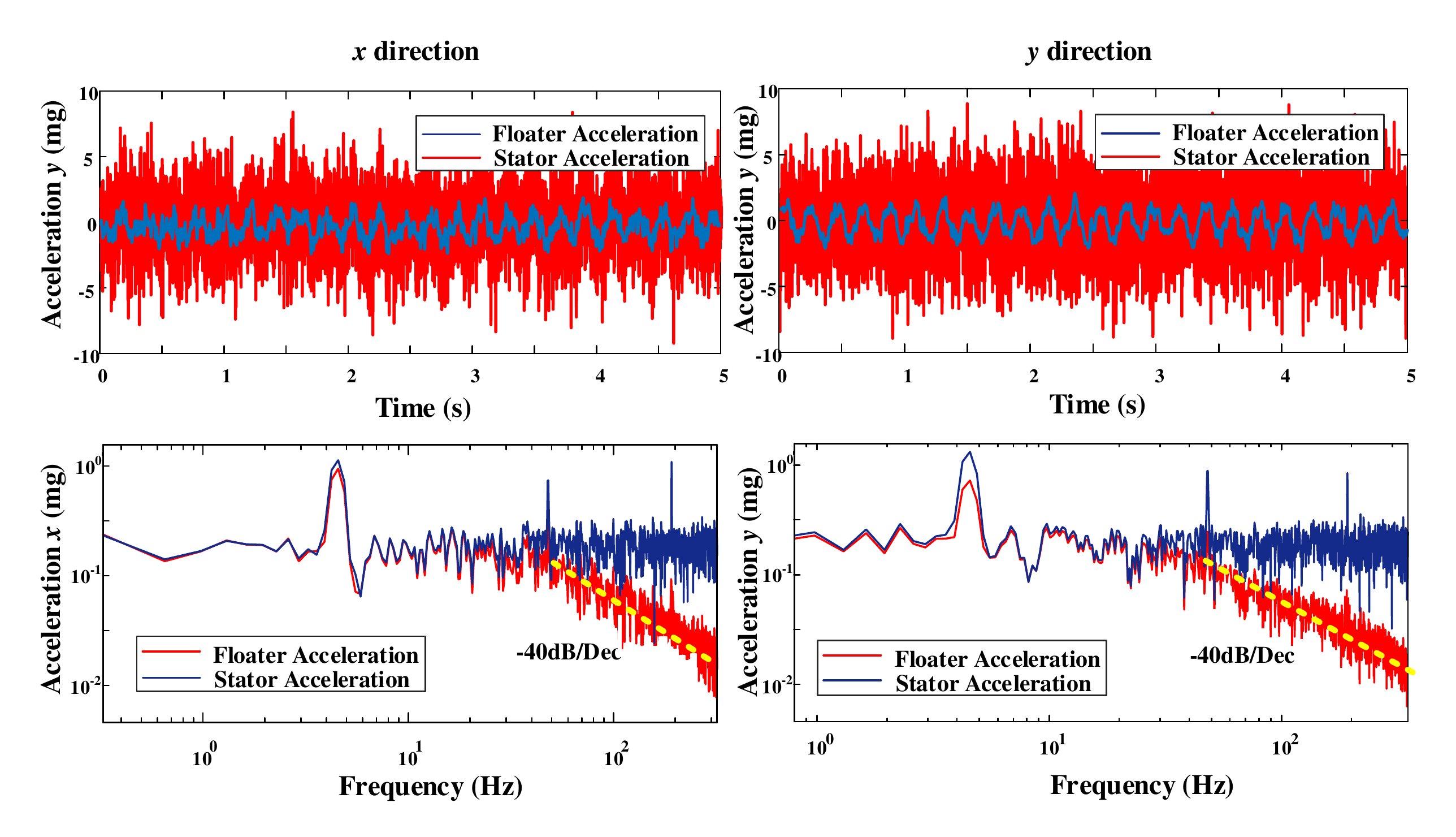}
\caption{Comparison of vibration level on stator and floater with HAFIMC control}
\label{fig:15}
\end{figure}

In contrast to IMC control, the adaptive feedforward part is added to deal with the periodical disturbance.
The experiment conditions are the same as the previous section.
The acceleration level of stator and floater are shown in Fig.\ref{fig:15}.
From the frequency domain part in Fig.\ref{fig:15}, one can see that, while ensuring system stability and suppression of random noise, the proposed HAFIMC controller has a notable inhibitory effect on periodical disturbances, which unquestionably confirm the effectiveness of the suggested control scheme.
For the $x$-axis, the rejection rate of the 50 Hz signal reaches -22.51 dB contrasting to -10.86 dB for IMC controller.
The rejection rate of the 200 Hz signal reaches -39.75 dB contrasting to -21.49 dB for IMC controller.
More detailed data is listed in the Table.\ref{table:3}.

According to Figs.\ref{fig:14}, Figs.\ref{fig:15} and Table.\ref{table:3}, it can be seen that the proposed HAFIMC controller achieve better vibration isolation performance than pure IMC control.
Based on the results mentioned above, the proposed HAFIMC controller does not only has the robustness features of IMC with the suppression for high frequency disturbances but also has additional suppression characteristics, which is introduced by adaptive feedforward control, for periodic disturbances.

\section{Conclusion}
A maglev vibration isolation platform (MVIP) is proposed and described in this paper as well as the corresponding modeling, coupling analysis and controller design.
To enable this system achieving payload-agnostic decoupling, we propose an SRBFNNI scheme which could approximate the unknown nonlinear inversion dynamic and linearize the system after payload redeployed.
To simultaneously deal with the two control problems introduced by the unique vibration isolation mechanism, this paper proposes a hybrid adaptive feedforward internal mode controller (HAFIMC) that could keep system stable while counteracting random and periodic disturbance.
Several experiments and simulations are carried out to verify the proposed scheme's effectiveness: Firstly, the SRBFNNI succeeds in decoupling and linearizing the nonlinear MIMO system automatically and precisely.
Secondly, the proposed HAFIMC maneuver MVIP achieves relative position maintenance and vibration isolation simultaneously.
Furthermore, the proposed hybrid scheme is superior to the pure feedback scheme in periodical vibration rejection without sacrificing system robustness.

\section*{Acknowledgements}
This study was supported by the National Natural Science Foundation of China (Grant No.51822502, 51475117), Foundation for Innovative Research Groups of the natural Science Foundation of China (Grant No. 51521003), the Fundamental Research Funds for the Central Universities (Grant No. HIT.BRETIV.201903).

 \appendix

\section{The Brief Derivation of Inversion System Scheme}
\label{Inversion System Scheme}
 
For a general system with $p$-dimensional input vector $\boldsymbol{u}(t) = [ u_{1}, \cdots , u_{p}]^{T}$, $p$-dimensional output vector $\boldsymbol{y}(t) = [ y_{1},\cdots , y_{p}]^{T}$,  $n$-dimensional states $\boldsymbol{x}(t) = \left[ x_{1}, \cdots , x_{n} \right]^{T}$ and initial state vector $\boldsymbol{x}(t_{0}) = \boldsymbol{x_{0}}$, the system equation can be expressed as:

\begin{equation}
\label{eq.A.1}
    \left\{\begin{array}{l}{\boldsymbol {\dot{x}}(t)=f(x_{1}, x_{2}, \cdots, x_{n}, u_{1}, u_{2}, \cdots, u_{p}), \quad \boldsymbol{x}\left(t_{0}\right)=\boldsymbol{x_{0}}} \\ {\boldsymbol{y}(t)=c(x_{1}, x_{2}, \cdots, x_{n},  u_{1}, u_{2}, \cdots, u_{p})}\end{array}\right.
\end{equation}
where $f$ and $c$ are the locally functions.
(\ref{eq.A.1}) can be regarded as an operator mapping the system input $\boldsymbol{u}(t)$ into the output $\boldsymbol{y}(t)$, which is expressed as

\begin{equation}
\label{eq.A.2}
    \boldsymbol{y}(\cdot)=\theta \cdot \left[\boldsymbol{x}_{0}, \boldsymbol{u}(\cdot)\right] \text { or } \boldsymbol{y}=\theta \cdot \boldsymbol{u}
\end{equation}

System inversion is capable of achieving the inverse mapping from the output $\boldsymbol{y}$ to the input $\boldsymbol{u}$.
If there exist a system $\Sigma_{\alpha}$ with mapping operator $\overline{\theta}_{\alpha} : \boldsymbol{\varphi} \rightarrow \boldsymbol{u}_{d}$, where $\boldsymbol{\varphi}(t)=\boldsymbol{y}_{d}^{(\alpha)}(t)$, that is, $\varphi$ is the $\boldsymbol{\alpha}$-th order derivative of $\boldsymbol{y_{d}}$
and satisfies 

\begin{equation}
\label{eq.A.3}
\theta \overline{\theta}_{\alpha}\left(\boldsymbol{y}_{d}^{(\alpha)}\right)=\theta \overline{\theta}_{\alpha} \varphi=\theta \boldsymbol{u}_{d}=\boldsymbol{y}_{d}  
\end{equation}

Then system $\Sigma_{\alpha}$ is called the $\alpha$-th order integral inversion system of (\ref{eq.A.1}) \cite{X.Dai2001}.
Cascading $\Sigma_{\alpha}$ with its original system, the mapping from the input $\boldsymbol{\varphi}$ to the output $\boldsymbol{y_{d}}$ will be identical with pure integration process, which can be formulated as 

\begin{equation}
\label{eq.A.4}
\boldsymbol{y}_{d} = \theta \boldsymbol{u}_{d} = 
\frac{1}{s^{\alpha}} \boldsymbol{y}_{d}^{(\alpha)}  
\end{equation}
where $s$ denotes the Laplace variable.
Then the combined system could be decoupled into several independent identical pseudo integrators.

 \section{The Derivation of adaptive feed-forward controller}
 \label{FXLMS}

Given the $W_{i}(z)$ is an N-order adaptive FIR filter whose impulse $W_{i}(n)$ can be denoted as

\begin{align}
    W_{i}(n) =\left[W_{i,0}(n),W_{i,1}(n), \dots,  W_{i,N-1}(n) \right]^{T}
\end{align} 

Then the output $U_{i}(n)$ could be calculated by the convolution of $X_{i}(n)$ and $W_{i}(n)$

\begin{equation}
U_{i}(n) = \sum_{j=0}^{N-1} W_{i, j}(n) X_{i}(n-j)= X_{i}^{T}(n)W_{i}(n)
\end{equation}

\noindent where the reference signal $X_{i}(n)$ comes from the measured acceleration on stator can be denoted as

\begin{equation}
X_{i}(n)= \left[{x_{i}(n)}, {x_{i}(n-1)}, {\dots}, {x_{i}(n-N+1)} \right]^{T}
\end{equation}

The secondary path $H_{i}(z)$ represents the discredited dynamics model that transmitting the control output $U_{i}(n)$ to the measuring point.
Its impulse coefficient can be expressed as

\begin{equation}
{H}_{i}(n)=\left[h_{i, 0}(n), {h_{i, 1}(n)}, {\dots}, {h_{i, N-1}(n)} \right]^{T}
\end{equation}

Supposing the estimated dynamics of secondary path $\hat{H}_{z}$ is equal to ${H}_{z}$, then the cancellation signal $Y_{i}(n)$ can be derived as

\begin{equation}
\begin{aligned}
Y_{i}(n) &= \sum_{j=0}^{N-1} h_{i, j}(n) u_{i}(n-j) = {U}_{i}^{T}({n}) {H}_{i}({n}) = W_{i}^{T}(n) X_{i}(n) H_{i}(n) \\ &= W_{i}^{T}(n) X_{i}^{'}(n)
\end{aligned}
\end{equation}

\noindent where $X_{i}^{'}(n) = X_{i}^{T}(n) \hat{H}_{i}(n)$ denote the aligned feed-forward signal.
The output of the floater, $E(n)$, which is also the measurement of accelerometer on floater, can be expressed in z-domain as

\begin{equation}
    E_{i}(z^{-1}) = D_{i}(z^{-1}) - Y_{i}(z^{-1}) = [P_{i}(z^{-1}) - H_{i}(z^{-1})W_{i}(z^{-1})]X_{i}(z^{-1})
\end{equation}
where $D_{i}(z^{-1})$ is the disturbance desired to be cancelled.
It is transmitted from the $X_{i}(z^{-1})$ through $P(z^{-1})$.

The aim of adaptive vibration control is to obtain an proper $W_{i}(n)$ to full-fill the equation below and leads to $e_{i}(z^{-1}) = 0$.

\begin{equation}
\label{eq.condition_of_FXLMS}
    W_{i}(z^{-1}) = H_{i}^{-1}(z^{-1}) P_{i}(z^{-1})
\end{equation}

The $W_{i}(z^{-1})$ is adaptively update by minimizing the expectation of $e_{i}^{2}(n)$ which can be derived as 

\begin{equation}
\begin{aligned}
    Minimize \quad E[e_{i}^{2}(n)] = E[(D_{i}(n) - Y_{i}(n))^{2}]
    =E[(X_{i}^{T}(n) P_{i}(n) - W_{i}^{T}(n) X_{i}(n) H_{i}(n))^{2}]
\end{aligned}
\end{equation}

To solve this problem, the normalized least mean squre algorithm, which is the so called NLMS, is utilized here to adaptively update the cofficients of controller $W_{i}(n)$.
The gradient of $E[e_{i}^{2}(n)]$ is

\begin{equation}
    \frac{\partial e_{i}^{2}(n)}{\partial {W}_{i}(n)} = \left[ 2e_{i}(n)\frac{\partial e_{i}^{2}(n)}{\partial {W}_{i,0}(n)}, 2e_{i}(n)\frac{\partial e_{i}^{2}(n)}{\partial {W}_{i,1}(n)}, \cdots, 2e_{i}(n)\frac{\partial e_{i}^{2}(n)}{\partial {W}_{i,N-1}(n)} \right] = -2e_{i}(n)X_{i}^{'}(n)
\end{equation}

Then the update rule of the coefficients of $W_{i}(n)$ can be given as

\begin{equation}
{W}_{i}(n+1)={W}_{i}(n)+2 \mu e_{i}(n) {X_{i}^{'}(n)}
\end{equation}



\bibliographystyle{model1-num-names}
\bibliography{Reference}







\end{document}